\documentclass[a4paper,11pt]{article}

\pdfoutput=1 

\usepackage{bbm}
\usepackage{jheppub} 
\usepackage[T1]{fontenc}
\usepackage{simplewick} 
\usepackage{makecell}	
\usepackage{tabu}		
\usepackage[makeroom]{cancel} 
\usepackage[normalem]{ulem} 
\usepackage[utf8]{inputenc}
\usepackage{amsmath,amsfonts,amsthm,amssymb} 
\usepackage{subcaption} 

\newcommand{\p}{\partial}
\newcommand{\eq}{&\quad}

\newcommand{\qRq}{\quad\Rightarrow\quad}
\newcommand{\rig}{\right.}
\newcommand{\lef}{\left.}

\newcommand{\para}{\parallel}

\newcommand{\eff}{\text{eff}}

\newcommand{\vol}{\text{vol}}

\newcommand{\mco}{\mathcal{O}}

\newcommand{\R}{\mathbb{R}}

\newcommand{\1}{\mathbbm{1}}

\newcommand{\hp}{{\hat{\phi}}}
\newcommand{\hs}{{\hat{\sigma}}}
\newcommand{\hx}{{\hat{x}}}

\newcommand{\al}{\alpha}
\newcommand{\be}{\beta}

\newcommand{\de}{\delta}
\newcommand{\e}{\epsilon}
\newcommand{\ph}{\phi}
\newcommand{\g}{\gamma}

\newcommand{\la}{\lambda}
\newcommand{\m}{\mu}

\newcommand{\s}{\sigma}

\newcommand{\De}{\Delta}

\newcommand{\G}{\Gamma}

\preprint{UUITP-09/23}

\title{\boldmath Fusion of conformal defects in interacting theories}

\author{Alexander Söderberg Rousu}

\affiliation{Department of Physics and Astronomy,
	Uppsala University,\\
	Box 516,
	SE-751 20 Uppsala,
	Sweden}

\emailAdd{alexander.soderberg.rousu@gmail.com}

\makeatletter
\gdef\@fpheader{}
\makeatother

\abstract{
	We study fusion of two scalar Wilson defects. We propose that fusion holds at a quantum level by showing that bare one-point functions stay invariant. This is an expected result as the path integral stays invariant under fusion of the two defects. The difference instead lies in renormalization of local quantities on the defects. Those on the fused defect takes into account UV divergences in the fusion limit when the two defects approach eachother, in addition to UV divergences in the coincident limit of defect-local fields and in the near defect limits of bulk-local fields. At the fixed point of the corresponding RG flow the two conformal defects have fused into a single conformal defect.
	
	Parts of this paper was first presented in my thesis \cite{SoderbergRousu:2023ucv}.
}

\begin{document} 
	
\newtheorem{defin}{Definition}
\newtheorem{thm}{Theorem}
\newtheorem{cor}{Corollary}
\newtheorem{pf}{Proof}
\newtheorem{nt}{Note}
\newtheorem{ex}{Example}
\newtheorem{ans}{Ansatz}
\newtheorem{que}{Question}
\newtheorem{ax}{Axiom}

\maketitle

\section{Introduction}

A defect is an extended object of dimension $p\geq 1$. E.g. a line or a surface. In this paper we study systems with two defects. The Poincar\'e or conformal symmetry in the bulk is broken in the same way as for one defect: each defect will be charged under an orthogonal group $SO(d - p)$ (with $p$ being the dimension of the defect), and a defect-local field with support on a defect is charged under $SO(p - 1, 1) \times SO(d - p)$ (assuming flat defects). Due to localization, each defect is only affected by the nearby bulk theory. 

Local characteristics, such as anomalous dimensions, $\be$-functions and \textit{operator product expansion} (OPE) coefficients, of the bulk theory are not affected by the defects (since the \textit{ultra-violet} (UV) divergences these quantities arise from are the coincident-limits of bulk fields). Likewise, the corresponding characteristics on each defect are not affected by other defects (since these defect quantities arise from their corresponding coincident-limits of defect-local fields and defect-limits of bulk-local fields).

There will be several new OPE's in play. In addition to the usual bulk-bulk OPE there is a defect-defect OPE (similar to the bulk-bulk OPE) on each defect and a \textit{defect operator product expansion} (DOE) for each defect. The DOE allows us to expand bulk-local fields in terms of defect-local ones \cite{diehl1986field, Billo:2016cpy}.

If the defects intersect, there is also one defect-intersection DOE for each defect and an intersection-intersection OPE. In the conformal case (when the theories on the intersection, both of the defects and the bulk are all conformal), these give rise to a conformal bootstrap equation for bulk one- and bulk-intersection two-point functions \cite{Antunes:2021qpy}.


In this paper we will consider two (parallel) \textit{scalar Wilson defects} (or pinning defects) separated by a distance $2\,R$. These conformal defects are given by
\begin{equation} \label{Scalar Wilson loop}
\begin{aligned}
D = \exp\left( i\,h\int_{\R^p}d^px_\para \hp \right) \ ,
\end{aligned}
\end{equation}
where $h$ describes a magnetic field along the defect\footnote{This is seen from the equation of motion, where $h$ will act as a source term along the defect.} and hatted operator denote those localized to the defect. From a technical point of view, $h$ can be treated as a coupling constant of finite size localized on the defect \cite{Cuomo:2021kfm}.  See \cite{Cuomo:2021rkm, Cuomo:2022xgw, Rodriguez-Gomez:2022gbz, Rodriguez-Gomez:2022xwm, Rodriguez-Gomez:2022gif, Aharony:2022ntz, Bolla:2023zny, Pannell:2023pwz} for recent development on these defects. The dimension of the defect is $p = 1$ (a line) if $d= 4 - \e$, and $p = 2$ (a surface) if $d = 6 - \e$. Both of these two models have their $O(N)$-symmetry explicitly broken by the scalar Wilson defect. See \cite{Giombi:2022vnz} for a similar defect in a fermionic QFT.

In \cite{Soderberg:2021kne}, \textit{fusion} of two scalar Wilson defects was studied in the four dimensional free theory. In the limit $R\rightarrow 0$ it was found that the two defects can be described by a single which does not preserve the conformal symmetry
\begin{equation} \label{fusion}
\begin{aligned}
D_f = \exp\left( -2\,h \sum_{n \geq 0} \frac{R^{2\,n}}{(2\,n)!}\int_{\R}dx_\para\p_R^{2\,n}\hp(x_\para) \right) \ .
\end{aligned}
\end{equation}
One way to understand this statement is that the distance, $R$, between the two defects is a scale of the theory, and thus has to be preserved after the fusion. This scale then enters in the interactions on $D_f$, making them dimensionfull. In turn, this makes the fused defect action non-conformal. 

In the language of fusion categories \cite{etingof2005fusion, bartels2019fusion, douglas2020dualizable}, this is an example when there is only when fused defect (with the OPE coefficient being one)
\begin{equation}
\begin{aligned}
D(-R) D(+R) = D_f(0) \ .
\end{aligned}
\end{equation}
Unlike \cite{etingof2005fusion, bartels2019fusion, douglas2020dualizable}, the defects were fused in \cite{Soderberg:2021kne} without using super or topological symmetry.

In this paper we study fusion of two scalar Wilson defects \eqref{Scalar Wilson loop} in interacting theories, and find the \textit{renormalization group} (RG) flow of the interactions on $D_f$ \eqref{fusion}. As expected, the dimensionfull couplings will not have well-defined \textit{fixed points} (f.p.'s). This means that after we have fused the defects, we can turn on interactions in the bulk and find a f.p. for $D_f$ where we have restored the conformal symmetry.

We will mostly consider a model with cubic bulk-interactions in $d = 6 - \e$. In Sec. \ref{Sec: RG} we study the one-point function of bulk fields in the presence of the two defects and find the RG flow for the defect couplings. This is a slight generalization of the corresponding results in \cite{Rodriguez-Gomez:2022gbz}, and we use the more traditional way of calculating Feynman diagrams \cite{Cuomo:2021kfm} assuming the bulk interactions are small w.r.t. those on the defects. In particular, we find that the couplings on the two defects are not affected by each other, which is what we expect since their corresponding $\be$-functions measure UV divergences in their respective defect-limit of bulk-local fields as well as UV divergences in the coincident-limit of defect-local fields on the corresponding defect.

In Sec. \ref{Sec: Fusion} we improve the results from \cite{Soderberg:2021kne}, which concerns fusion of scalar Wilson defects \eqref{Scalar Wilson loop} in $d = 4$ free theories. In the free theory we generalize this result to hold for any $d$. Specifying to the real-valued f.p.'s of the defects in $d = 6 - \e$, we compare the bare one-point function in the presence of $D_+$ and $D_-$ with that near $D_f$ (upto second order in the bulk couplings). We find that they are exactly the same, and there are no modifications needed to $D_f$. The underlying reason for this is that the path integral for $D_+$ and $D_-$ is the same as that for $D_f$. We check that this is indeed the case for line defects in $d = 4 - \e$ with a quartic bulk-interaction as well.



The difference between the theory with two defects and that with the fused defect lies in renormalization of the theory. Diagrams with bulk vertices connecting the two defects have logarithmic divergences in the \textit{fusion-limit} (as the distance between the defects goes to zero). Such divergences are absorbed in the bare coupling constants on the fused defect, giving us different $\be$-functions and renormalized correlators. So in addition to UV divergences in the coincident-limit of defect-local fields and in the defect-limit of bulk-local fields, the $\be$-functions on the fused defect also take into account UV divergences in the fusion-limit of the two defects.

\section{Renormalization group fixed points} \label{Sec: RG}

Let us first introduce the main model we consider. In the bulk we have
\begin{equation} \label{Bulk action}
\begin{aligned}
S = \int_{\R^d}d^dx\left(\frac{(\p_\m\ph^i)^2}{2} + \frac{(\p_\m\s)^2}{2} + \frac{g_1}{2}\s(\ph^i)^2 + \frac{g_2}{3!}\s^3\right) \ ,
\end{aligned}
\end{equation}
where $d = 6 - \e$ and $i\in\{1, ..., N\}$. The scalars $\ph^i$ are invariant under $O(N)$. We consider two parallel surface defects, $D_\pm$, of dimension $p = 2$, spanned along $\hx_\para^a ,\ a\in\{1, 2\}$. They are separated by a distance $2\,R$, $R \equiv |R_i|$, in the orthogonal directions $\hx_\perp^i,\ i\in\{1, ..., d - p\}$
\begin{equation} \label{Defects}
\begin{aligned}
D_\pm = \exp\left( -\int_{\R^p}d^px\big[h^\ph_\pm\hp^{i_\pm}(x_\pm) + h^\s_\pm\hs(x_\pm)\big] \right) \ .
\end{aligned}
\end{equation}
Here $x_\pm \equiv x_a\hx_\para^a \pm R_i\hx_\perp^i$ and $h^\ph_\pm$, $h^\s_\pm$ are couplings (or magnetic fields) of finite size localized on the respective defects. Due to their $\ph^{i_\pm}$-interaction, the $O(N)$-symmetry of the model is broken down to $O(N - 2)$ by the defects (in the case when $i_+ = i_-$ the symmetry is broken down to $O(N - 1)$). This is an explicit symmetry breaking caused by the defect interactions, and thus differs from e.g. the extraordinary p.t. near a boundary (which is a spontaneous symmetry breaking \cite{domb2000phase, PhysRevB.47.5841, Shpot:2019iwk}).

The effective action is given by
\begin{equation} \label{eff S}
\begin{aligned}
S_\eff = S + \sum_{\pm}\log D_\pm \ .
\end{aligned}
\end{equation}
Since the $\be$-functions for the bulk couplings arise from divergences in the coincident-limit of the bulk fields, they are not affected by the defect couplings. This means that we can borrow these results from the bulk theory \eqref{Bulk action} without the defects \cite{10.1143/PTP.54.1828, Fei:2014yja}
\begin{equation}
\begin{aligned}
\be_1 &= -\frac{\e}{2}g_1 + \frac{(N - 8)g_1^3 - 12g_1^2g_2 + g_1\,g_2^2}{12(4\,\pi)^3} + \mco(g^4) \ , \\
\be_2 &= -\frac{\e}{2}g_2 - \frac{4\,N\,g_1^3 - N\,g_1^2g_2 + 3g_2^3}{4(4\,\pi)^3} + \mco(g^4) \ .
\end{aligned}
\end{equation}
In the case when $N = 0$ and the $\ph^i$-fields are not present, there is a negative sign in front of the $g_2^3$-term in $\be_2$. Due to this we find no real-valued RG f.p. 

\begin{figure} 
	\centering
	\includegraphics[width=0.5\textwidth]{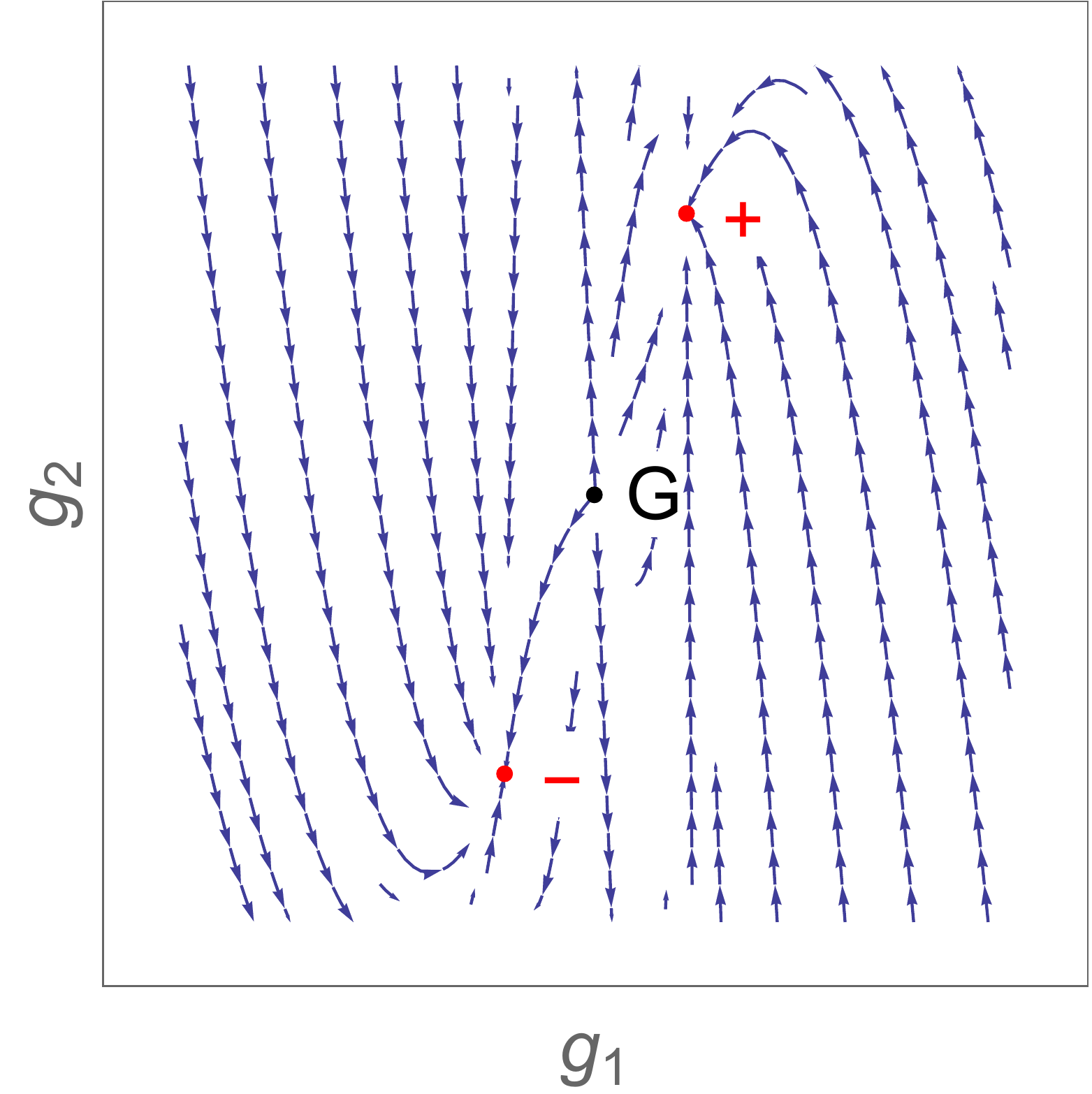}
	\caption{The RG flow of the cubic $O(N)$-model \eqref{Bulk action} at large $N$ near six dimensions. The black dot ($G$) is the trivial Gaussian f.p. and the two red dots ($\pm$) are the attractive f.p.'s at \eqref{Bulk couplings}.}
	\label{Fig: RGphi3}
\end{figure}

By including the $O(N)$-scalars we can expand in large $N \gg 1$ 
\begin{equation} \label{Bulk beta}
\begin{aligned}
\be_1 &= -\frac{\e}{2}g_1 + \frac{N\,g_1^3}{12(4\,\pi)^3} + \mco(g^4) \ , \\
\be_2 &= -\frac{\e}{2}g_2 - \frac{4\,N\,g_1^3 - N\,g_1^2g_2}{4(4\,\pi)^3} + \mco(g^4) \ .
\end{aligned}
\end{equation}
Setting these $\be$-functions to zero yields in addition to a Gaussian f.p., two non-trivial, real-valued f.p.'s\footnote{In particular, this result is valid for $N \geq 1039$ \cite{Fei:2014yja}}
\begin{equation} \label{Bulk couplings}
\begin{aligned}
g_2^* = 6\, g_1^* + \mco\left(\e, \frac{1}{N}\right) \ , \quad g_1^* = \pm\sqrt{\frac{6(4\,\pi)^3\e}{N}} + \mco\left(\e, \frac{1}{N}\right) \ .
\end{aligned}
\end{equation}
Note that these f.p.'s go as $\sqrt{\e}$, which differ from the \textit{Wilson-Fisher} (WF) f.p. in $d = 4 - \e$.  The RG flow is depicted in Fig. \ref{Fig: RGphi3}.

We will proceed with finding the f.p.'s of the defect-interactions in \eqref{Defects}. The corresponding $\be$-functions measure divergence in the respective near distance limits. This means that e.g. the $\be$-functions on $D_+$ does not depend on the interactions on $D_-$. In turn this tells us that the defect $\be$-functions are the same on the two defects, and it can be found from the theory with only one defect. We will calculate Feynman diagrams in the theory with both defects (considering small bulk-interactions) to show that this is indeed the case. 

\subsection{Free theory} \label{Sec: 6d free}

Correlators in the presence of the two defects \eqref{Defects} are found by expanding $D_\pm$ in its interactions and then applying Wick's theorem. This was done for a single insertion of a bulk field in \cite{Soderberg:2021kne}. In general, it gives us
\begin{equation}
\begin{aligned}
\langle D_+D_-...\rangle &= \langle D_+\rangle\langle D_-\rangle\langle D_+D_-\rangle \left( \sum_{\pm}\langle D_\pm...\rangle_N + \de\langle D_+D_-...\rangle_N \right) \ .
\end{aligned}
\end{equation}
Here the dots represent any combination of operators. $\langle D_\pm\rangle$ describes self-interactions on $D_\pm$, and $\langle D_+D_-\rangle$ is a non-perturbative (w.r.t. $R$) Casimir effect between the defects. See Fig. 2 in \cite{Soderberg:2021kne} for a diagrammatic representation of these correlators. 


For the purposes of this Section we are not interested in $\langle D_\pm\rangle$ and $\langle D_+D_-\rangle$. Thus we normalize correlators in the following way\footnote{Likewise if we consider $n$ defects, $\{D_i\}_{i = 1}^n$, then we use the normalization
	\begin{equation*}
	\begin{aligned}
	\langle D_1...D_n...\rangle_N \equiv \frac{\langle D_1...D_n...\rangle}{\langle D_1\rangle...\langle D_n\rangle\langle D_1...D_n\rangle} = \sum_{i = 1}^n\langle D_i...\rangle_N + \de\langle D_1...D_n...\rangle_N \ ,
	\end{aligned}
	\end{equation*}
	where $\de\langle D_1...D_n...\rangle_N$ contains Feynman diagrams connecting two or more defects.}
\begin{equation} \label{Normalization}
\begin{aligned}
\langle D_+D_-...\rangle_N \equiv \frac{\langle D_+D_-...\rangle}{\langle D_+\rangle\langle D_-\rangle\langle D_+D_-\rangle} = \sum_{\pm}\langle D_\pm...\rangle_N + \de\langle D_+D_-...\rangle_N \ .
\end{aligned}
\end{equation}
The remaining three correlators, $\langle D_\pm...\rangle_N$ and $\de\langle D_+D_-...\rangle_N$, can be found using standard Feynman diagrams techniques. $\langle D_\pm...\rangle$ is the one-point function in the presence of the single defect $D_\pm$, and $\de\langle D_+D_-...\rangle$ is the sum of Feynman diagrams connecting the two defects. We will see examples of $\de\langle D_+D_-...\rangle$-diagrams later in when we take into account the bulk-interactions.

The one-point function of $\ph^i$ and $\s$ in the presence of the two defects $D_\pm$ are given by the third Feynman diagram in Fig. 2 of \cite{Soderberg:2021kne}
\begin{equation} \label{One-pt fcn 1} 
\begin{aligned}
\langle D_+D_-\ph^i(x)\rangle_N &= \sum_{\pm}\langle D_\pm\ph^i(x)\rangle_N \ , \quad &\langle D_\pm\ph^i(x)\rangle_N &= -h_\pm^\ph\de^{ii_\pm}K_\pm(x) \ , \\
\langle D_+D_-\s(x)\rangle_N &= \sum_{\pm}\langle D_\pm\s(x)\rangle_N \ , \quad &\langle D_\pm\s(x)\rangle_N &= -h_\pm^\s K_\pm(x) \ ,
\end{aligned}
\end{equation}
where the integral $K_\pm$, which is the corresponding integral from \cite{Soderberg:2021kne}, is given by
\begin{equation} \label{int K 0}
\begin{aligned}
K_\pm(x) = \int_{\R^p}d^pz\langle\s(x)\s(z_\pm)\rangle\big|_{h_\pm^\ph, h_\pm^\s = 0} \ .
\end{aligned}
\end{equation}
The integrand is the same as the connected part of $\langle D_+D_-\s(x)\s(y)\rangle_N$. It is not affected by the defects interactions, and is thus the massless scalar correlator found from the Klein-Gordon equation 
\begin{equation} \label{2 pt fcn}
\begin{aligned}
\langle\s(x)\s(y)\rangle\big|_{h_\pm^\ph, h_\pm^\s = 0} = \langle D_+D_-\s(x)\s(y)\rangle_N^\text{conn} = \frac{A_d}{|x - y|^{2\,\De_\ph}} \ .
\end{aligned}
\end{equation}
The constant $A_d$ is given by
\begin{equation} \label{Solid angle}
\begin{aligned}
A_d = \frac{1}{(d - 2)S_d} \ , \quad S_d = \frac{2\,\G_{\frac{d}{2}}}{\pi^{\frac{d}{2}}} \ .
\end{aligned}
\end{equation}
Here $S_d$ is the solid angle and $\G_x \equiv \G(x)$ is a shorthand notation for the Gamma function. The integrals $K_\pm$ are thus given by
\begin{equation}
\begin{aligned}
K_\pm(x) &= A_dI^p_{\De_\ph}(0, x_\para, x_\perp \mp R) \ ,
\end{aligned}
\end{equation}
which are written in terms of the master integral (given in terms of a modified Bessel function of the second kind)
\begin{equation} \label{Master int 1}
\begin{aligned}
I^n_\De(k, w, z^2) &= \int_{\R^n}d^nx \frac{e^{i\,k\,x}}{[(x - w)^2 + z^2]^{\De}} \\
&= \frac{\pi^{\frac{n}{2}}}{2^{\De - \frac{n}{2} - 1}\G_\De}e^{i\,k\,w}\left( \frac{|k|}{|z|} \right)^{\De - \frac{n}{2}} K_{\De - \frac{n}{2}}(|k|\,z) \\
&= \left\{\begin{array}{l l}
\displaystyle{\frac{\pi^{\frac{n}{2}}\G_{\frac{n}{2} - \De}}{2^{2\De - n}\G_\De}\frac{e^{i\,k\,w}}{|k|^{2\,\De - n}}} \ , \quad &\text{if $z = 0$} \ , \\
\frac{}{} \\
\displaystyle{\frac{\pi^{\frac{n}{2}}\G_{\De - \frac{n}{2}}}{\G_\De}\frac{1}{|z|^{2\,\De - n}}} \ , \quad &\text{if $k = 0$} \ .
\end{array}\right.
\end{aligned}
\end{equation}
This yields
\begin{equation} \label{int K}
\begin{aligned}
K_\pm(x) &= \frac{A_d\pi^{\frac{p}{2}}\G_{\De_\ph - \frac{p}{2}}}{\G_{\De_\ph}} \frac{1}{|x_\perp \mp R|^{2\,\De_\ph - p}} \ .
\end{aligned}
\end{equation}
In the interacting theory we will find it useful to Fourier transform w.r.t. the normal distances, $s_\perp^\pm \equiv x_\perp \mp R$, to the defects
\begin{equation} \label{1-pt prop}
\begin{aligned}
\prod_{c = \pm}\int_{\R^{d - p}}ds_\perp^c e^{i\,k_\perp^c s_\perp^c} K_\pm(x) &= \frac{A_d\pi^{\frac{p}{2}}\G_{\De_\ph - \frac{p}{2}}}{\G_{\De_\ph}}\de(k_\pm)I^{d - p}_{\De_\ph - \frac{p}{2}}(k_\perp^\mp, 0, 0) \\
&= \frac{\de(k_\pm)}{k_\mp^2} \ , \quad \text{(exactly).}
\end{aligned}
\end{equation}
The momenta $k_\pm$ is that flowing between the bulk field and the defect $D_\pm$. It describes how momenta is being absorbed/emitted by the two defects. The Dirac $\de$-function tell us that the momenta is only affected by one of the defects in the free theory.

Note that the one-point functions \eqref{One-pt fcn 1} are the forms we expect a one-point function to have from conformal symmetry \cite{Billo:2016cpy},\footnote{The defects are placed at the orthogonal coordinates $\pm R_i$, hence a shift in the denominator.} from which we can read off the DOE coefficients
\begin{equation}
\begin{aligned}
\m^{\ph^i}{}_{\1_\pm} = -h_\pm^\ph\de^{ii_\pm}\frac{\G_{\De_\ph - 1}}{4\,\pi^{\De_\ph}} \ , \quad \m^{\s}{}_{\1_\pm} = -h_\pm^\s\frac{\G_{\De_\ph - 1}}{4\,\pi^{\De_\ph}} \ ,
\end{aligned}
\end{equation}
where the $\1_\pm$ subscript denotes the identity exchange on the respective defect.

\subsection{Interacting theory}

We will now proceed to the interacting theory, and find the $\be$-functions of the defect couplings as well as the corresponding RG f.p.'s.

\begin{figure}
	\begin{subfigure}{.5\textwidth}
		\centering
		\includegraphics[width=.5\linewidth]{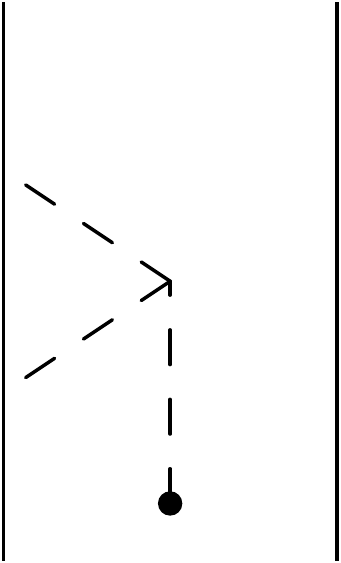}
	\end{subfigure}%
	\begin{subfigure}{.5\textwidth}
		\centering
		\includegraphics[width=.5\linewidth]{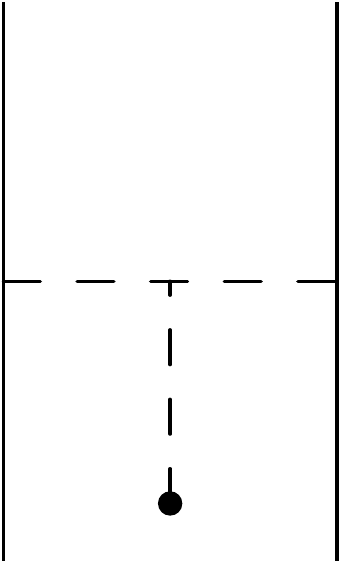}
	\end{subfigure}
	\caption{The two Feynman diagrams that contribute to the one-point functions of $\ph^i$ and $\s$. The dot is the external bulk point, the dotted lines are either $\ph - \ph$ or $\s - \s$ correlators and the solid lines are the two surface defects.}
	\label{fig: diag}
\end{figure}

The one-point functions at $\mco(g)$ are given by the two Feynman diagrams in Fig. \ref{fig: diag}. If a diagram contains $n$ defect points of the same field, we have to divide the symmetry factor with a factor $n!$ to avoid overcounting (which is seen from the integration of the defect points). We find
\begin{equation} \label{Int one-pt fcn}
\begin{aligned}
\langle D_\pm\ph^i(x)\rangle^{(1)}_N &= 2\left( -\frac{g_1}{2} \right) (-h^\ph_\pm)(-h^\s_\pm)\de^{ii_\pm}L^\pm_\pm(x) \ , \\
\de\langle D_+D_-\ph^i(x)\rangle^{(1)}_N &= 2\left( -\frac{g_1}{2} \right) \sum_{a = \pm}(-h^\ph_a)(-h^\s_{-a})\de^{ii_a}L^+_-(x) \ , \\
\end{aligned}
\end{equation}
and
\begin{equation}\label{Int one-pt fcn 2}
\begin{aligned}
\langle D_\pm\s(x)\rangle^{(1)}_N &= \left( \frac{2}{2} \left( -\frac{g_1}{2} \right)(-h^\ph_\pm)(-h^\ph_\pm)\de^{i_\pm i_\pm}  + \rig \\
\eq \lef + \frac{3!}{2} \left( -\frac{g_2}{3!} \right)(-h^\s_\pm)(-h^\s_\pm) \right) L^\pm_\pm(x) \ , \\
\de\langle D_+D_-\s(x)\rangle^{(1)}_N &= \left( 2 \left( -\frac{g_1}{2} \right)(-h^\ph_+)(-h^\ph_{-})\de^{i_+ i_{-}} +  \rig \\
\eq \lef + 3! \left( -\frac{g_2}{3!} \right)(-h^\s_+)(-h^\s_{-}) \right) L^+_{-}(x) \ .
\end{aligned}
\end{equation}
Here $L^a_b$, with $a,b = \pm$, is the following integral
\begin{equation} \label{Int Lab}
\begin{aligned}
L^a_b(x) &\equiv \int_{\R^d}d^dz \left.\langle \ph(x)\ph(z)\rangle\right|_{h = 0}K_a(z)K_b(z) \\
&= A_d\int_{\R^{d - p}}d^{d - p}z_\perp K_a(z)K_b(z)I^p_{\De_\ph}(0, x_\para, (z_\perp - x_\perp)^2) \\
&= \frac{A_d^3\pi^{\frac{3\,p}{2}}\G_{\De_\ph - \frac{p}{2}}^3}{\G_{\De_\ph}^3}\int_{\R^{d - p}}\frac{d^{d - p}z_\perp}{(|z_\perp|\,|z_\perp + x_\perp - a\,R|\,|z_\perp + x_\perp - b\,R|)^{2\,\De_\ph - p}} \ ,
\end{aligned}
\end{equation}
where in the last step we shifted $z_\perp \rightarrow z_\perp + x_\perp$. When $a = b$ this integral can be solved using the following master integral
\begin{equation} \label{Master int 2}
\begin{aligned}
J^n_{a,b}(z) &\equiv \int_{\R^n}\frac{d^nx}{|x|^{2\,a}|x - z|^{2\,b}} \\
&= \frac{\G_{a + b}}{\G_a\G_b}\int_{0}^{1}du(1 - u)^{a - 1}u^{b - 1}\int_{\R^n}\frac{d^nx}{(x^2 + u(1-u)z^2)^{a + b}} \\
&= \frac{\pi^{\frac{n}{2}}\G_{a + b - \frac{n}{2}}\G_{\frac{n}{2} - a}\G_{\frac{n}{2} - b}}{\G_a\G_b\G_{n - a - b}}\frac{1}{|z|^{2(a + b) - n}} \ .
\end{aligned}
\end{equation}
It gives us
\begin{equation} \label{int L p p}
\begin{aligned}
L^\pm_\pm(x) &= A_d^3\pi^{\frac{3\,p}{2}}\frac{\G_{\De_\ph - \frac{p}{2}}^3}{\G_{\De_\ph}^3}J^{d - p}_{\De_\ph - \frac{p}{2}, 2\,\De_\ph - p}(x_\perp \mp R) \\
&= \frac{\G_{\frac{d + p}{2} - 2\,\De_\ph}\G_{\frac{d}{2} - \De_\ph}\G_{\De_\ph - \frac{p}{2}}^2\G_{3\,\De_\ph - \frac{d}{2} + p}}{\G_{d + \frac{p}{2} - 3\,\De_\ph}\G_{\De_\ph}^3\G_{2\,\De_\ph - p}} \frac{A_d^3\pi^{\frac{3\,p}{2}}}{|x_\perp \mp R|^{6\,\De_\ph - d - 2\,p}} \ .
\end{aligned}
\end{equation}
We find it easier to study the UV divergences of the integral $L^+_-$ in momentum space, where we Fourier transform w.r.t. $s_\perp^\pm$
\begin{equation}
\begin{aligned}
M^a_b(k_\perp^\pm) &\equiv \prod_{c = \pm}\int_{\R^{d - p}}d^{d - p}s_\perp^c e^{i\, k_\perp^c s_\perp^c} L^a_b(x) \ .
\end{aligned}
\end{equation}
This integral can then be performed using only the master integral \eqref{Master int 1}.
\begin{equation} \label{Int M ++}
\begin{aligned}
M^\pm_\pm(k_\perp^\pm) &= \frac{A_d^3\pi^{\frac{3\,p}{2}}\G_{\De_\ph - \frac{p}{2}}^3}{\G_{\De_\ph}^3} \de(k_\perp^\pm) \int_{\R^{d - p}}\frac{d^{d - p}z_\perp}{|z_\perp|^{2\,\De_\ph - p}} I^p_{2\,\De_\ph - p}(k_\perp^\mp, -z_\perp, 0) \\
&= \frac{A_d^32^{d + p - 4\,\De_\ph}\pi^{\frac{d}{2} + p}\G_{\frac{d + p}{2} - 2\,\De_\ph}\G_{\De_\ph - \frac{p}{2}}^3}{\G_{\De_\ph}^3\G_{2\,\De_\ph - p}|k_\perp^\mp|^{\frac{d + p}{2} - 2\,\De_\ph}} \de(k_\perp^\pm) I^{d - p}_{\De_\ph - \frac{p}{2}}(-k_\perp^\mp, 0, 0) \\
&= \frac{\de(k_\perp^\pm)}{8\pi^2(k_\perp^\mp)^2} \left( \frac{1}{\e} - \log|k_\perp^\mp| + \mathcal{A} \right) \ ,
\end{aligned}
\end{equation}
\begin{equation} \label{Int M +-}
\begin{aligned}
M^+_-(k_\perp^\pm) &= \frac{A_d^3\pi^{\frac{3\,p}{2}}\G_{\De_\ph - \frac{p}{2}}^3}{\G_{\De_\ph}^3} \int_{\R^{d - p}}\frac{d^{d - p}z_\perp}{|z_\perp|^{2\,\De_\ph - p}} \prod_{a = \pm} I^p_{\De_\ph - \frac{p}{2}}(k_\perp^a, -z_\perp, 0) \\
&= \frac{A_d^34^{d - 2\,\De_\ph}\pi^{d + \frac{p}{2}}\G_{\frac{d}{2} - \De_\ph}^2\G_{\De_\ph - \frac{p}{2}}}{\G_{\De_\ph}^3|k_\perp^+|^{d - 2\,\De_\ph}|k_\perp^-|^{d - 2\,\De_\ph}} I^{d - p}_{\De_\ph - \frac{p}{2}}(-k_\perp^+ - k_\perp^-, 0, 0) \\
&= \frac{1}{(k_\perp^+)^2(k_\perp^-)^2(k_\perp^+ + k_\perp^-)^2} \ , \quad \text{(exactly).}
\end{aligned}
\end{equation}
Here $\mathcal{A}$ is the following constant 
\begin{equation}
\begin{aligned}
\mathcal{A} &= \log\left( \frac{2\sqrt{\pi}}{e^{\frac{\g_E}{2} - 1}} \right) \ ,
\end{aligned}
\end{equation}
which can be absorbed in the coupling constants (by defining \textit{minimal subtraction} (MS) scheme couplings) without affecting the RG flow. Thus we will not care about it.

Note that the Feynman diagram $M^+_-$ (in $\de\langle D_+D_-\ph^i(k_\pm)\rangle^{(1)}_N$) connecting the two defects is convergent. So only the diagrams $M^\pm_\pm$ (in $\langle D_\pm\ph^i(k_\pm)\rangle^{(1)}_N$), which are affected by one of the defects, are divergent. This means that in the renormalization procedure, the couplings on $D_+$ are not affected by those on $D_-$ (and vice versa). This is the expected result since the bare couplings on $D_+$ should capture UV divergences in the coincident-limit of defect-local fields in addition to divergences in the limit as bulk-local fields approach $D_+$.\footnote{At higher orders in the bulk couplings, there are divergent diagrams in $\de\langle D_+D_-\ph^i\rangle$. However, the divergences in these diagrams are taken care of by renormalization of local quantities at lower orders in the bulk couplings. One such example is the first diagram in Fig. \ref{fig: g2Fusion}.}


To find the bare defect couplings we add the free theory correlators \eqref{One-pt fcn 1} to those at first order in the coupling constants \eqref{Int one-pt fcn}. We then make the following ansatz for the bare coupling constants
\begin{equation} \label{D+D- bare}
\begin{aligned}
h_\pm^\ph &= \m^{\frac{\e}{2}} \tilde{h}_\pm^\ph \left( 1 + a_\pm^\ph\frac{\tilde{h}_\pm^\s\tilde{g}_1}{\e} \right) \ , \quad h_\pm^\s = \m^{\frac{\e}{2}} \left( \tilde{h}_\pm^\s + b_\pm^\ph\frac{(\tilde{h}_\pm^\ph)^2\tilde{g}_1}{\e} + b_\pm^\s\frac{(\tilde{h}_\pm^\s)^2\tilde{g}_2}{\e} \right) \ ,
\end{aligned}
\end{equation}
where the constants $a_\pm^\ph, b_\pm^\ph$ and $b_\pm^\s$ are tuned s.t. that the $\e$-poles in the correlators vanish. Coupling constants with a tilde are renormalized ones (dimensionless), and $\m$ is the RG scale. By expanding the correlators in the bulk couplings and in $\e$ we find (by matching powers of $k_\pm$)
\begin{equation}
\begin{aligned}
b_\pm^\ph = b_\pm^\s = \frac{a_\pm^\ph}{2} \ , \quad a_\pm^\ph &= -\frac{1}{8\,\pi^2} \ .
\end{aligned}
\end{equation}
From which we find the $\be$-functions (by differentiating $\log\,h^\ph_\pm$, $\log\,h^\s_\pm$ w.r.t. $\log\m$)\footnote{Here we used that the bulk $\be$-functions are given by \eqref{Bulk beta}. See e.g. App. B of \cite{Prochazka:2020vog} for details on this.}
\begin{equation}
\begin{aligned}
\be_\pm^\ph &= -\frac{\e}{2}\tilde{h}_\pm^\ph - \frac{\tilde{h}_\pm^\ph\tilde{h}_\pm^\s\tilde{g}_1}{8\,\pi^2\e} \ , \quad \be_\pm^\s &= -\frac{\e}{2}\tilde{h}_\pm^\s - \frac{(\tilde{h}_\pm^\ph)^2\tilde{g}_1}{16\,\pi^2\e} - \frac{(\tilde{h}_\pm^\s)^2\tilde{g}_2}{16\,\pi^2\e} \ .
\end{aligned}
\end{equation}
Setting these to zero gives us a Gaussian f.p. where both defect couplings are zero.\footnote{The defect couplings can be zero while those in the bulk \eqref{Bulk couplings} are not.} We also find the following non-trivial ones
\begin{equation}
\begin{aligned}
\bigg( (h_\pm^\ph)^*, (h_\pm^\s)^* \bigg) \in \left\{ \left(0, -\frac{8\,\pi^2\e}{g_2^*}\right), \left(\pm 4\,\pi^2\e\frac{\sqrt{2\,g_1^* - g_2^*}}{(g_1^*)^{\frac{3}{2}}}, -\frac{4\,\pi^2\e}{g_1^*}\right) \right\} \ .
\end{aligned}
\end{equation}
The first one is the same as that found in \cite{Rodriguez-Gomez:2022gbz}. The bulk couplings are tuned to their respective f.p.'s \eqref{Bulk couplings}, where we find four complex f.p.'s
\begin{equation} \label{RG fp 0}
\begin{aligned}
\bigg( (h_\pm^\ph)^*, (h_\pm^\s)^* \bigg) = \left( \pm i\sqrt{\frac{\pi\, N\, \e}{6}}, \pm\frac{1}{2}\sqrt{\frac{\pi\, N \,\e}{6}}\right) \ ,
\end{aligned}
\end{equation}
and two real-valued f.p.'s where only $h_\pm^\s$ is non-trivial
\begin{equation} \label{RG fp}
\begin{aligned}
\bigg( (h_\pm^\ph)^*, (h_\pm^\s)^* \bigg) = (0, h^*) \ , \quad h^* = \mp\frac{1}{6}\sqrt{\frac{\pi\, N\,\e}{6}} \ .
\end{aligned}
\end{equation}
The sign of $h^*$ is opposite to the bulk-couplings at their f.p. \eqref{Bulk couplings}. If we restrict ourselves to real-valued f.p.'s then the $\ph^{i_\pm}$-term on the defects \eqref{Defects} vanish
\begin{equation} \label{Defects 2}
\begin{aligned}
D_\pm = \exp\left( -h^*\int_{\R^p}d^px\,\hs(x_\pm) \right) \ .
\end{aligned}
\end{equation}
Note that since $N\gg 1$, none of the f.p.'s (\ref{RG fp 0}, \ref{RG fp}) have to be small. 

By studying the derivative of $\be^\s_\pm$ we can check whether the real-valued f.p. is attractive or not
\begin{equation}
\begin{aligned}
\p_{\tilde{h}^\s_\pm}\be^\s_\pm\bigg|_{\tilde{h}^\ph_\pm = 0, \ \tilde{h}^\s_\pm = h^*} = \frac{\e}{2} \ ,
\end{aligned}
\end{equation}
which does not depend on the sign of $h^*$ at \eqref{RG fp}. Since this is positive, the f.p.'s at \eqref{RG fp} are minima of the defect $\s$-coupling and are thus attractive. 

The one-point functions of $\ph^i$ are trivial at this f.p. (restoring $O(N)$-symmetry), while those for $\s$ can be resummed in $\e$
\begin{equation}
\begin{aligned}
\langle D_+D_-\s(k_\perp^\pm)\rangle_N &= \sum_{a = \pm} \langle D_a\s(k_\perp^\pm)\rangle_N + \de\langle D_+D_-\s(k_\perp^\pm)\rangle_N + \mco(g^2) \ ,
\end{aligned}
\end{equation}
\begin{equation} \label{<s>}
\begin{aligned}
\langle D_\pm\s(k_\perp^\pm)\rangle_N &= -h^*\frac{\de(k_\perp^\pm)}{(k_\perp^\mp)^{2 - \frac{\e}{2}}} \ , \\
\de\langle D_+D_-\s(k_\perp^\pm)\rangle_N &= -\frac{(h^*)^2g_2^*}{(k_\perp^+)^2(k_\perp^-)^2(k_\perp^+ + k_\perp^-)^2} \ .
\end{aligned}
\end{equation}
Note that the RG scale has completely vanished at the f.p. We have
\begin{equation}
\begin{aligned}
(h^*)^2g_2^* = \pm\frac{\pi^\frac{5}{2}}{\sqrt{N}}\left( \frac{2\,\e}{3} \right)^\frac{3}{2} \ ,
\end{aligned}
\end{equation}
which is at a subleading order in $N$. This means that $\de\langle D_+D_-\s(k_\perp^\pm)\rangle_N$ is small compared to $\langle D_\pm\s(k_\perp^\pm)\rangle_N$.

In Euclidean space we find
\begin{equation} \label{Euclidean space}
\begin{aligned}
\langle D_\pm\s(x)\rangle_N &= \prod_{c = \pm}\int_{\R^{d - p}}\frac{d^{d - p}k_\perp^c}{(2\,\pi)^{d - p}} e^{-i\,k_\perp^cs_\perp^c} \langle D_\pm\s(k_\perp^\pm)\rangle_N  \\
&= -\frac{h^*}{(2\pi)^{d - p}} I^{d - p}_{1 - \frac{\e}{4}}(-s_\perp^\pm, 0, 0) + \mco(\e^\frac{3}{2}) = -\frac{h^*}{(2\pi)^{2 - \frac{\e}{2}}|s_\perp^\pm|^{2 - \frac{\e}{2}}} \ ,
\end{aligned}
\end{equation}
which agrees with the free theory result at $\mco(\sqrt{\e})$, and has the correct scaling dimension of $\s$. From this we can also read off the DOE coefficients
\begin{equation}
\begin{aligned}
\m^{\s}{}_{\1_\pm} = -\frac{h^*}{(2\pi)^{2 - \frac{\e}{2}}} + \mco(g^2) \ .
\end{aligned}
\end{equation}

\subsection{Order $g^2$} \label{Sec: fp}

\begin{figure}
	\centering
	\includegraphics[width=.5\linewidth]{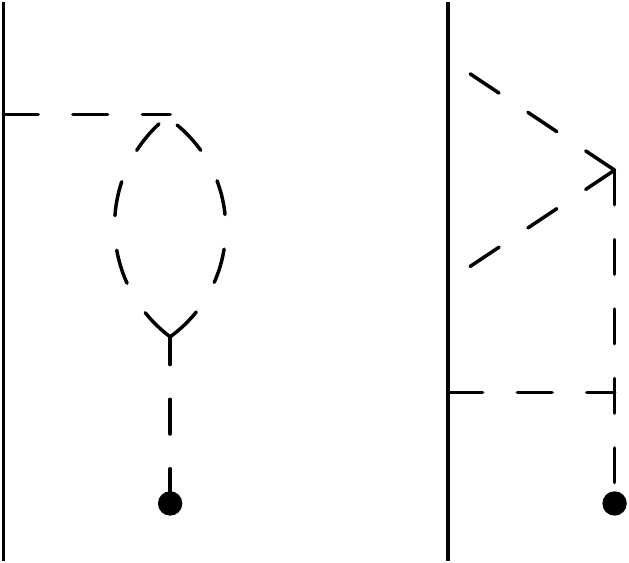}
	\caption{The two Feynman diagrams at $\mco(g^2)$ in $\langle D_+\s\rangle$. In the bulk-loop of the first diagram there are either $\ph$- or $\s$-internal fields.}
	\label{fig: g2}
\end{figure}

Before we study fusion, let us calculate the Feynman diagrams at $\mco(g^2)$, and find the defect f.p. upto $\mco(\e^{\frac{3}{2}})$. To do this we study the one-point function of $\s$ in the presence of one defect \eqref{Defects 2} with only a $\s$-interaction. At $\mco(g^2)$ it is given by
\begin{equation} \label{Ds 2}
\begin{aligned}
\langle D_+\s \rangle^{(2)} &= \left( 2\,N \left(-\frac{g_1}{2}\right)^2 + 3*3!\left(-\frac{g_2}{2}\right)^2 \right) (-h) A + \frac{(3!)^2}{2!} \left(-\frac{g_2}{2}\right)^2 (-h)^3 B \ ,
\end{aligned}
\end{equation}
where $A$ is the first diagram in Fig. \ref{fig: g2}, and $B$ the second
\begin{equation} \label{ints A and B 0}
\begin{aligned}
A &= \prod_{i = 1}^2\int_{\R^d}d^dz_i\left.\langle\s(x)\s(z_1)\rangle\right|_{h = 0}\left.\langle\s(z)\s(z_2)\rangle\right|_{h = 0}^2K_+(z_2) \ , \\
B &= \prod_{i = 1}^2\int_{\R^d}d^dz_i \left. \langle\s(x)\s(z_1)\rangle \right|_{h = 0} \left. \langle\s(z)\s(z_2)\rangle \right|_{h = 0} K_+(z_1) K_+(z_2)^2 \ .
\end{aligned}
\end{equation}
These integrals can be calculated using the master integrals (\ref{Master int 1}, \ref{Master int 2}). In order, we perform the following steps:
\begin{enumerate}
	\item Integrate over the parallel vertex coordinates: $z_\para^i \in \R^p$.
	\item Integrate over $z_\perp \in \R^{d - p}$.
	\item Integrate over $w_\perp \in \R^{d - p}$.
	\item For simplicity, Fourier transform w.r.t. $s_\perp \equiv x_\perp - R \in \R^{d - p}$ and express the diagram in terms of the orthogonal momenta $k_\perp$. We denote the Fourier transform of $A$, $B$ with $\tilde{A}$, $\tilde{B}$ respectively.
	\item Neglect constant-terms at $\mco(\e^0)$ which can be absorbed in the coupling constants in the MS scheme (for simplicity).
\end{enumerate}
This gives us
\begin{equation} \label{ints A and B}
\begin{aligned}
\tilde{A} &= \pi^{\frac{3\,d}{2}}A_d^4 \frac{ \G_{\frac{d}{2} - 2\,\De_\ph} \G_{\frac{d}{2} - \De_\ph} }{\G_{2\,\De_\ph} \G_{\De_\ph}^2} \left( \frac{2}{k_\perp^-} \right)^{3\,d - 8\,\De_\ph} = -\frac{1}{192\,\pi^3k_\perp^2} \left( \frac{1}{\e} - \frac{\log(k_\perp^2)}{2} + \mco(\e) \right) \ , \\
\tilde{B} &= \pi^{\frac{3\,d}{2} + p}A_d^5 \frac{ \G_{\frac{d + p}{2} - 4\,\De_\ph} \G_{d + p - 4\,\De_\ph} \G_{3\,\De_\ph - \frac{d}{2} - p} \G_{\frac{d}{2} - \De_\ph}^2 \G_{\frac{d}{2} - \De_\ph}^2 }{ \G_{d + \frac{p}{2} - 3\,\De_\ph} \G_{2\,\De_\ph - p} \G_{4\,\De_\ph - \frac{d + 3\,p}{2}} \G_{\De_\ph}^5 } \left( \frac{2}{k_\perp^-} \right)^{3\,d + 2(p - 5\,\De_\ph)} \\
&= \frac{1}{128\,\pi^4k_\perp^2} \left( \frac{1}{\e^2} + \frac{ \mathcal{A}^{(2)} - \log(k_\perp^2) }{\e} + \left( \frac{\log(k_\perp^2)}{2} \right)^2 - 2\,\mathcal{A}^{(2)}\,\log(k_\perp^2) \right) \ .
\end{aligned}
\end{equation}
Here $\tilde{A}$ captures the contribution to the bulk anomalous dimension of $\s$, and $\mathcal{A}^{(2)}$ is the following constant
\begin{equation}
\begin{aligned}
\mathcal{A}^{(2)} &= \log\left( \frac{4\,\pi}{e^{\g_E - \frac{5}{2}}} \right) \ .
\end{aligned}
\end{equation}
The $\frac{\mathcal{A}^{(2)}}{\e}$-term in $B$ \eqref{ints A and B} will cancel due to the $\mco(\e)$-term from $M^+_+$ \eqref{Int M ++} when we renormalize the full one-point function of $\s$. This serves as a good sanity check on our result. Upto $\mco(g^2)$, $\langle\s\rangle$ it is given by
\begin{equation} \label{one pt fcn O(g2)}
\begin{aligned}
\langle\s(k)\rangle &= -\frac{h}{k_\perp^2} \left( 1 + \frac{g_2h\csc\left(\frac{\pi\,\e}{2}\right)}{4^{3 - \e}\pi^{\frac{1 - \e}{2}}\G_{\frac{3 - \e}{2}}|k_\perp|^\e} - \frac{(Ng_1^2 + g_2^2)\csc\left(\frac{\pi\,\e}{2}\right)}{4^{5 - \e}\pi^{\frac{3 - \e}{2}}\G_{\frac{5 - \e}{2}}|k_\perp|^\e} + \rig \\
\eq \lef + \frac{g_2^2h^2\e\csc(\pi\,\e)\G_{-\frac{\e}{2}}^2}{2^{11 - 3\,\e}\pi^{\frac{5}{2} - \e}\G_{2 - \frac{3\,\e}{2}}\G_{\frac{3 - \e}{2}}|k_\perp|^\e} + \mco(g^3) \right) \ .
\end{aligned}
\end{equation}
At $\mco(g^2)$ the field $\s$ receives an anomalous dimension \cite{Fei:2014yja}. Thus we also have to introduce a $Z$-factor for this field
\begin{equation} \label{Z-factor 1}
\begin{aligned}
\s = \sqrt{Z}\tilde{\s} \qRq \tilde{\s} = \frac{\s}{\sqrt{Z}} \ , \quad \sqrt{Z} = 1 + z\frac{N\,g_1^2 + g_2^2}{\e} + \mco(g^3) \ .
\end{aligned}
\end{equation}
In this $Z$-factor we have a coefficient $z$ which we can find from the bulk theory without a defect. Note that when we compute $\langle\s\rangle$ at \eqref{int K 0} in the free theory we integrate over the two-point function \eqref{2 pt fcn} of $\s$ in the presence of no defect ($h = 0$). This means that we need to include an extra factor of $\sqrt{Z}$ every time the integral $K_+$ appear in the Feynman diagrams in (\ref{Int Lab}, \ref{ints A and B 0}) \cite{Pannell:2023pwz}.\footnote{We are grateful to Diego Rodriguez-Gomez for a discussion on this.} Technically, this means that we should divide every bare defect coupling, $h$, with $\sqrt{Z}$ in the one-point function of the renormalized field $\tilde{\s}$
\begin{equation} \label{renorm s}
\begin{aligned}
\langle\tilde{\s}(k)\rangle &= -\frac{h}{Z\,k_\perp^2} \left( 1 + \frac{g_2h\csc\left(\frac{\pi\,\e}{2}\right)}{4^{3 - \e}\pi^{\frac{1 - \e}{2}}\G_{\frac{3 - \e}{2}}\sqrt{Z}\,|k_\perp|^\e} - \frac{(Ng_1^2 + g_2^2)\csc\left(\frac{\pi\,\e}{2}\right)}{4^{5 - \e}\pi^{\frac{3 - \e}{2}}\G_{\frac{5 - \e}{2}}|k_\perp|^\e} + \rig \\
\eq \lef + \frac{g_2^2h^2\e\csc(\pi\,\e)\G_{-\frac{\e}{2}}^2}{2^{11 - 3\,\e}\pi^{\frac{5}{2} - \e}\G_{2 - \frac{3\,\e}{2}}\G_{\frac{3 - \e}{2}}Z\,|k_\perp|^\e} + \mco(g^3) \right) \ .
\end{aligned}
\end{equation}
At this order in the bulk couplings, only the $Z$-factor at $\mco(g^0)$ will contribute. 

To renormalize $\langle\tilde{\s}(k)\rangle$ we make the following ansatz for the bare couplings
\begin{equation} \label{bare def}
\begin{aligned}
g_i &= \m^\frac{\e}{2}\tilde{g}_i + \mco(g^2) \ , \quad i\in\{1, 2\} \ , \\
h &= \m^{\frac{\e}{2}}\tilde{h} \left( 1 - \frac{\tilde{g}_2\tilde{h}}{16\,\pi^2\e} + a\frac{N\,g_1^2 + g_2^2}{\e} + \tilde{g}_2^2\tilde{h}^2 \left( \frac{b_2}{\e^2} + \frac{b_1}{\e} \right) + \mco(g^3) \right) \ ,
\end{aligned}
\end{equation}
where $a$, $b_1$ and $b_2$ are three coefficients to be fixed by cancelling the poles in $\e$. By expanding in the bulk couplings and then $\e$, we are able to cancel the poles with
\begin{equation}
\begin{aligned}
a = \frac{1}{384\,\pi^3} + z \ , \quad b_2 = \frac{1}{256\,\pi^4} = \left( -\frac{1}{16\,\pi^2} \right)^2 \ , \quad b_1 = -\frac{1}{512\,\pi^4} \ .
\end{aligned}
\end{equation}
Note that $b_2$-term in the bare coupling \eqref{bare def} is exactly twice the coefficient in front of the $\mco(g_2h^2)$-term. This serves as a consistency check on our result as it will cancel an $\e$-pole in our $\be$-function (which we will soon calculate).

As input from the bulk theory, the $Z$-factor coefficient, $z$, will precisely tune $a$ to zero
\begin{equation} \label{Z-factor 2}
\begin{aligned}
z = -\frac{1}{384\,\pi^3} \qRq a = 0 \ .
\end{aligned}
\end{equation}
This is another consistency check of our result, since if $a$ were to be non-zero, the f.p. \eqref{RG fp} from order $\mco(g)$ would get further corrections at $\mco(\sqrt{N\,\e})$ (due to the $\mco(Ng_1^2)$-term in $\langle\tilde{\s}\rangle$).

All and all, we find the bare defect coupling
\begin{equation} \label{bare def 2}
\begin{aligned}
h &= \m^{\frac{\e}{2}}\tilde{h} \left( 1 - \frac{\tilde{g}_2\tilde{h}}{16\,\pi^2\e} + \tilde{g}_2^2\tilde{h}^2 \left( \frac{1}{256\,\pi^4\e^2} + \frac{1}{512\,\pi^4\e} \right) + \mco(g^3) \right) \ ,
\end{aligned}
\end{equation}
and the one-point function (neglecting constants of $\mco(\e^0)$)
\begin{equation}
\begin{aligned}
\langle\tilde{\s}(k)\rangle &= -\frac{\tilde{h}}{k_\perp^2} \left( 1 - \frac{\tilde{g}_2\tilde{h}}{32\,\pi^2}\log\left( \frac{k_\perp^2}{\m^2} \right) - \frac{N\,\tilde{g}_1^2 + \tilde{g}_2^2}{768,\pi^3} \log\left( \frac{k_\perp^2}{\m^2} \right) + \rig \\
\eq\lef + \frac{\tilde{g}_2^2\tilde{h}^2}{1024\,\pi^4} \left( \log\left( \frac{k_\perp^2}{\m^2} \right) - \mathcal{B} \right)^2 + \mco(g^3) \right) \ , \\
\mathcal{B} &= \log\left( \frac{2\sqrt{\pi}}{e^{\frac{\g_E - 3}{2}}} \right) \ .
\end{aligned}
\end{equation}
As another sanity check we find that all logarithms are dimensionless.

Finally, from \eqref{bare def 2} we find the $\be$-function for the defect coupling 
\begin{equation}
\begin{aligned}
\be_h = -\frac{\tilde{h}\,\e}{2} - \frac{\tilde{g}_2\tilde{h}^2}{16\,\pi^2} - \frac{\tilde{g}_2^2\tilde{h}^3}{256\,\pi^4} + \mco(g^3) \ , 
\end{aligned}
\end{equation}
which has the perturbative f.p.
\begin{equation} \label{Def fp}
\begin{aligned}
h^* = -8\,\pi^2 \left( \frac{1}{\tilde{g}_2} \mp \sqrt{\frac{1 - 2\,\e}{\tilde{g}_2^2}} \right) \ .
\end{aligned}
\end{equation}
The sign in front of the square root is opposite to that in the bulk f.p. \eqref{Bulk couplings}, which we write out here again to higher orders of $\e$ and $N^{-1}$ \cite{Fei:2014yja}
\begin{equation*}
\begin{aligned}
g_1^* &= \pm\sqrt{\frac{6(4\,\pi)^3\e}{N}} \left[ 1 + \frac{22}{N} + \frac{726}{N^2} - \frac{326,180}{N^3} - \frac{349,658,330}{N^4} + \mco\left(\frac{1}{N^5}\right) + \rig \\
\eq \lef - \frac{\e}{N} \left[ \frac{155}{6} + \frac{1,705}{N} - \frac{912,545}{N^2} - \frac{3,590,574,890}{N^3} + \mco\left( \frac{1}{N^4} \right) \right] + \mco\left(\frac{\e^3}{N^3}\right) \right] \ ,
\end{aligned}
\end{equation*}
\begin{equation*}
\begin{aligned}
g_2^* &= \pm\sqrt{\frac{6(4\,\pi)^3\e}{N}} \left[ 6 \left[ 1 + \frac{162}{N} + \frac{68,760}{N^2} + \frac{41,224,420}{N^3} + \frac{28,762,554,870}{N^4} + \mco\left(\frac{1}{N^5}\right) \right] + \rig \\
\eq \lef - \frac{\e}{N} \left[ \frac{215}{2} + \frac{86,335}{N} - \frac{75,722,265}{N^2} - \frac{69,633,402,510}{N^3} + \mco\left( \frac{1}{N^4} \right) \right] + \mco\left(\frac{\e^3}{N^3}\right) \right] \ .
\end{aligned}
\end{equation*}
If we now expand the defect f.p. \eqref{Def fp} in small $\e$ and large $N$ we find the defect f.p.
\begin{equation*}
\begin{aligned}
h^* &= \mp\frac{1}{6}\sqrt{\frac{\pi\, N\,\e}{6}} \left( 1 - \frac{162}{N} - \frac{45,522}{N^2} - \frac{23,195,764}{N^3} - \frac{15,402,417,210}{N^4} + \mco\left(\frac{1}{N^5}\right) + \rig \\
\eq \lef + \e \left[ \frac{1}{2} - \frac{757}{2\,N} - \frac{76,061}{N^2} - \frac{9,386,189}{2\,N^3} - \frac{4,845,204,490}{N^4} + \mco\left( \frac{1}{N^5} \right) \right] + \mco(\e^3) \right) \ .
\end{aligned}
\end{equation*}

\section{Fusion} \label{Sec: Fusion}

Let us now fuse the two defects \eqref{Defects 2}. This can be done by Taylor expanding $D_+$ and $D_-$  w.r.t. each component of $R_i$ (remember that the two defects are placed at $\pm R_i$ along the orthogonal coordinates)
\begin{equation}
\begin{aligned}
D_\pm = \exp\left( -h \prod_{i = 1}^{d - p} \sum_{n_i \geq 0} \frac{(\pm 1)^{n_i}R_i^{n_i}}{n_i!}\lim\limits_{R_i\rightarrow 0}\p_i^{n_i} \int_{\R^p}d^px\,\s(x_+) \right) \ .
\end{aligned}
\end{equation}
Adding the exponents gives us the fused defect
\begin{equation} \label{Df}
\begin{aligned}
D_f = D_+D_- = \exp\left( -2\,h \prod_{i = 1}^{d - p} \sum_{n_i \geq 0} \frac{R_i^{2\,n_i}}{(2\,n_i)!}\int_{\R^p}d^px\,\p_i^{2\,n_i}\hs(x_+) \right) \ .
\end{aligned}
\end{equation}
This is the multivariate version of the result in \cite{Soderberg:2021kne}. Since this is just a Taylor expansion, we find the path integral, which generates all of the correlators, to be the same for $D_+D_-$ as for $D_f$ (see Sec. 3 of \cite{Soderberg:2021kne} for a proof on this). Note that the entire tower of terms w.r.t. $R_i$ has to be kept to find the same path integral. $R_i$ should be treated as a distance scale of the theory, and thus we keep it even after fusion of the two defects.

Let us also mention that two straight parallel lines are conformally equivalent to two concentric circles. This means that above fusion is also true for two concentric circular Wilson lines. Although not commented upon, this was seen in \cite{Soderberg:2021kne}. I.e. its eq.'s (2.21) and (5.4) are the same. 


Since the path integral is the same for $D_+D_-$ and $D_f$, we expect the fusion \eqref{Df} to hold even in an interacting theory. In the rest of this paper we will perform several consistency checks to see that this indeed the case. Firstly, we will show that the defect correlators without any field insertions (the normalization factors in \eqref{Normalization}) are the same upto $\mco(g)$: $\langle D_+\rangle\langle D_-\rangle\langle D_+D_-\rangle = \langle D_f\rangle$. Then we will show that the expansion of $\langle D_+D_-\s\rangle_N$ in $R$ is the same as $\langle D_f\s\rangle_N$ upto $\mco(g^2)$ (before renormalization of the couplings). 

To simplify the calculations, we will choose a coordinate system s.t. $R$ is one-dimensional. In addition, we let the normal coordinate of the external field (when we study one-point functions) be one-dimensional as well
\begin{equation} \label{one dim R}
\begin{aligned}
R^i = R\,\de^{i1} \ , \quad R > 0 \ , \quad x_\perp^i = x_\perp\,\de^{i1} \ .
\end{aligned}
\end{equation}

\subsection{Normalization factor}

We will start by calculating the normalization factors. In the free theory, the logarithm of these correlators are given by the two first Feynman diagrams in Fig. 2 of \cite{Soderberg:2021kne}
\begin{equation}
\begin{aligned}
\log\langle D_\pm\rangle &= \frac{h^2}{2} \int_{\R^p}d^px\int_{\R^p}d^py \left.\langle\s(x_\pm)\s(y_\pm)\rangle\right|_{h = 0} = \frac{A_dh^2}{2}\vol(\R^p)\int_{\R^p}\frac{d^px}{(x^2)^{\De_\ph}} \ ,
\end{aligned}
\end{equation}
\begin{equation}
\begin{aligned}
\log\langle D_+D_-\rangle &= h^2 \int_{\R^p}d^px\int_{\R^p}d^py \left.\langle\s(x_+)\s(y_-)\rangle\right|_{h = 0} \\
&= A_dh^2\vol(\R^p)\int_{\R^p}\frac{d^px}{(x^2 + 4\,R^2)^{\De_\ph}} \\
&= A_dh^2\vol(\R^p)\sum_{n\geq 0}\binom{-\De_\ph}{n}(2\,R)^{2\,n}\int_{\R^p}\frac{d^px}{(x^2)^{\De_\ph + n}} \ ,
\end{aligned}
\end{equation}
where we do not perform the last (divergent) integral over $x$. Together they give
\begin{equation}
\begin{aligned}
\log(\langle D_+\rangle\langle D_-\rangle\langle D_+D_-\rangle) &= A_dh^2\vol(\R^p) \left( 2\int_{\R^p}\frac{d^px}{(x^2)^{\De_\ph}} + \rig \\
\eq\lef + \sum_{n\geq 1}\binom{-\De_\ph}{n}(2\,R)^{2\,n}\int_{\R^p}\frac{d^px}{(x^2)^{\De_\ph + n}} \right) \ .
\end{aligned}
\end{equation}
For $D_f$ we have a single diagram similar to the first one in Fig. 2 of \cite{Soderberg:2021kne}
\begin{equation}
\begin{aligned}
\log\langle D_f\rangle &= \frac{(2\,h)^2}{2!} \sum_{m_1, m_2 \geq 0} \frac{R^{2(m_1 + m_2)}}{(2\,m_1)!(2\,m_2)!}\int_{\R^p}d^px_\para\int_{\R^p}d^py_\para\times \\
\eq\times \lim\limits_{R', R'' \rightarrow 0}\p_{R'}^{2\,m_1}\p_{R''}^{2\,m_1}\left.\langle\s(x\hx_\para + R'\hx_\perp^1)\s(y\hx_\para + R''\hx_\perp^1)\rangle\right|_{h = 0} \\
&= 2\,A_dh^2\vol(\R^p) \sum_{n\geq 0}\sum_{m = 0}^{n}\frac{R^{2n}}{(2\,m)!(2\,n - 2\,m)!}\int_{\R^p}d^px_\para \times \\
\eq\times \lim\limits_{R', R'' \rightarrow 0}\p_{R'}^{2\,m}\p_{R''}^{2(n - m)}\frac{1}{(x^2 + (R' - R'')^2)^{\De_\ph + n}} \ .
\end{aligned}
\end{equation}
This is exactly $\log(\langle D_+\rangle\langle D_-\rangle\langle D_+D_-\rangle)$.

\begin{figure}
	\begin{subfigure}{.5\textwidth}
		\centering
		\includegraphics[width=.5\linewidth]{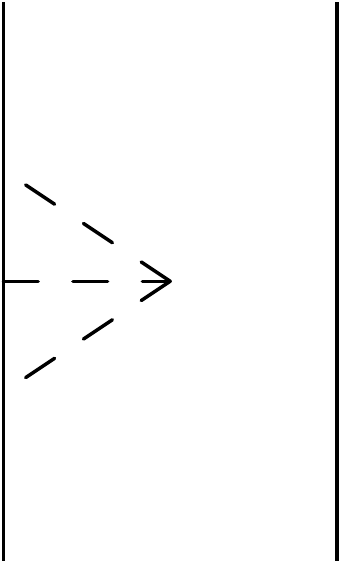}
	\end{subfigure}%
	\begin{subfigure}{.5\textwidth}
		\centering
		\includegraphics[width=.5\linewidth]{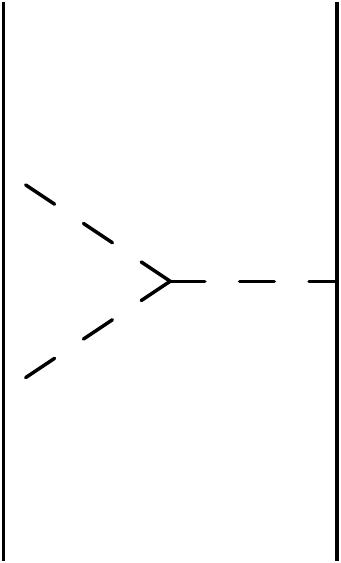}
	\end{subfigure}
	\caption{The two Feynman diagrams that contribute to the normalization factor at $\mco(g)$.}
	\label{fig: norm}
\end{figure}

\begin{figure}
	\centering
	\includegraphics[width=.13\linewidth]{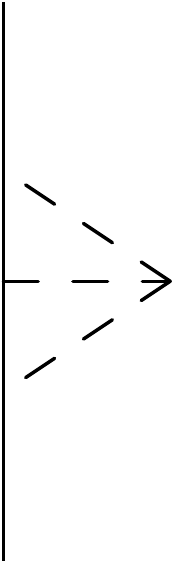}
	\caption{The single Feynman diagram that contribute to the normalization factor for the fused defect at $\mco(g)$.}
	\label{fig: norm Df}
\end{figure}

Let us now turn on the interactions and study these correlators at $\mco(g)$. For $D_\pm$ we have the two Feynman diagrams in Fig. \ref{fig: norm}
\begin{equation}
\begin{aligned}
\log\langle D_\pm\rangle^{(1)} &= \frac{3!}{3!} \left( -\frac{g_2}{3!} \right) (-h)^3 A^\pm_\pm \ , \\
\log\langle D_+D_-\rangle^{(1)} &= \frac{3!}{2!} \left( -\frac{g_2}{3!} \right) (-h)^3 \left(A^+_- + A^-_+\right) \ ,
\end{aligned}
\end{equation}
which gives us the full normalization factor
\begin{equation} \label{log DD}
\begin{aligned}
\log(\langle D_+\rangle\langle D_-\rangle\langle D_+D_-\rangle)^{(1)} &= \frac{g_2h^3}{2} \left( \frac{A^+_+ + A^-_-}{3!} + A^+_- + A^-_+ \right) \ .
\end{aligned}
\end{equation}
This is given in terms of the following integral
\begin{equation}
\begin{aligned}
A^a_b &= \int_{\R^d}d^dzK_a(z)^2K_b(z) \\
&= \frac{A_d^3\pi^{\frac{3\,p}{2}}\vol(\R^p)\G_{\De_\ph - \frac{p}{2}}^3}{\G_{\De_\ph}^3} \int_{\R^{d - p}}\frac{d^{d - p}z_\perp}{|z_\perp - a\,R|^{2\,\De_\ph - p}|z_\perp - b\,R|^{\De_\ph - \frac{p}{2}}} \ .
\end{aligned}
\end{equation}
Here we only integrated over the parallel part of the vertex ($z_\para \in \R^p$).
The normalization of $D_f$ is given by the single Feynman diagram in Fig. \ref{fig: norm Df}
\begin{equation}
\begin{aligned}
\log\langle D_f\rangle^{(1)} &= \frac{3!}{3!} \left( -\frac{g_2}{3!} \right) (-2\,h)^3 \int_{\R^d}d^dz \prod_{i = 1}^3 \sum_{m_i \geq 0} \frac{R^{2\,m_i}}{(2\,m_i)!}\int_{\R^p}d^px_i \times \\
\eq \times \lim\limits_{R_i\rightarrow 0} \p_{R_i}^{2\,m_i}\left.\langle\s(z)\s(x_i\hx_{\para} + R_i\hx_\perp^1)\rangle\right|_{h = 0} \ .
\end{aligned}
\end{equation}
Performing the integration over the parallel coordinates, and differentiating
\begin{equation} \label{b const}
\begin{aligned}
\lim\limits_{R'\rightarrow 0}\p_{R'}^{2\,m}[(a - R')^2]^{-\De_\ph + \frac{p}{2}} &= (2\,m)!\,b_m |a|^{-2\,\De_\ph + p - 2\,m} \ , \quad b_m \equiv \frac{( 2\,\De_\ph + p )_{2\,m}}{(2\,m)!} \ ,
\end{aligned}
\end{equation}
gives us
\begin{equation}
\begin{aligned}
\log\langle D_f\rangle^{(1)} &= 4\,A_d^3\pi^{\frac{3\,p}{2}}g_2h^3\vol(\R^p)\frac{\G_{\De_\ph - \frac{p}{2}}^3}{\G_{\De_\ph}^3}\sum_{n\geq 0} c_nR^{2\,n} \int_{\R^{d - p}}\frac{d^{d - p}z_\perp}{|z_\perp|^{3(2\,\De_\ph - p) + 2\,n}} \ .
\end{aligned}
\end{equation}
This is expressed in terms of the constant
\begin{equation}
\begin{aligned}
c_n &= \sum_{m = 0}^n\sum_{m' = 0}^m b_{n - m}b_{m - m'}b_{m'} \\
&= \sum_{m = 0}^n \frac{\G_{d - 4 + 2\,m}\G_{d - 4 + 2(n - m)}}{\G_{d - 4}^2\G_{2\,m + 1}\G_{2(n - m) + 1}} \times \\
\eq\times {}_4F_3\left( \frac{d - 2}{2}, \frac{d - 3}{2}, \frac{1}{2} - m, -m; \frac{1}{2}, \frac{5 - d}{2} - m, 3 - \frac{d}{2} - m; 1  \right) \ .
\end{aligned}
\end{equation}
By expanding \eqref{log DD} in $R$ we find perfect agreement with $\log\langle D_f\rangle^{(1)}$. Thus we have shown that the normalization factor is the same upto $\mco(g)$ (as expected)
\begin{equation}
\begin{aligned}
\langle D_f\rangle &= \langle D_+\rangle\langle D_-\rangle\langle D_+D_-\rangle \ .
\end{aligned}
\end{equation}

\subsection{One-point function} \label{Sec: one-pt fcn}

Let us now check that fusion also holds for the one-point function of $\s$. In the free theory we have
\begin{equation} 
\begin{aligned}
\langle D_f\s(x)\rangle^{(0)}_N &= -2\,A_d\pi^\frac{p}{2}h^2g_2\frac{\G_{\De_\ph - \frac{p}{2}}}{\G_{\De_\ph}} \sum_{n\geq 0} \frac{(2\,\De_\ph - p)_{2\,n}}{(2\,n)!}\frac{R^{2\,n}}{x_\perp^{2\,\De_\ph - p + 2\,n}} \\
&= -\frac{h^2g_2}{2\,\pi^2} \sum_{n\geq 1}(2\,n + 3)\frac{R^{2\,n}}{x_\perp^{2(n + 1)}} + \mco(\e) = \langle D_+D_-\s(x)\rangle^{(0)}_N \ ,
\end{aligned}
\end{equation}
which is in perfect agreement with $\langle D_+D_-\s(x)\rangle^{(0)}_N$ in \eqref{One-pt fcn 1}. 

At $\mco(g)$ we need the full $\langle D_+D_-\s\rangle^{(1)}_N$ in Euclidean space \eqref{Int one-pt fcn 2}. We already have $L^\pm_\pm$ \eqref{int L p p}, and are thus left to find 
\begin{equation}
\begin{aligned}
\de\langle D_+D_-\s(x)\rangle^{(1)}_N &= -(h^*)^2g_2^*L^+_- \ .
\end{aligned}
\end{equation}
We know from its Fourier transform \eqref{Int M +-} that $L^+_-$ is free of UV divergences. So we are free to set $\e = 0$ before integration over $z_\perp$ in \eqref{Int Lab}
\begin{equation} \label{L + -}
\begin{aligned}
L^+_- &= \int_{\R^4}\frac{d^4z_\perp}{64\,\pi^6}\frac{1}{z_\perp^2(z_\perp + s_\perp^+)^2(z_\perp + s_\perp^-)^2} \ .
\end{aligned}
\end{equation}
This integral has been done in the amplitude literature \cite{Chavez:2012kn}. Its a rather lengthy expression for general $R$, but by specifying to one dimensional $x_\perp$ and $R$ \eqref{one dim R} it simplifies to\footnote{This integral can also be done using Feynman parametrization. Then the integrals over the Feynman parameters simplify greatly in the case of \eqref{one dim R}.}
\begin{equation} 
\begin{aligned}
L^+_- &= \frac{1}{64\,\pi^4R} \left( \frac{1}{s_+}\log\left( \frac{|s_-|}{2\,R} \right) + \frac{1}{s_-}\log\left( \frac{|s_+|}{2\,R} \right) \right) \ .
\end{aligned}
\end{equation}
The full $\langle D_+D_-\s\rangle^{(1)}_N$ is thus
\begin{equation*}
\begin{aligned}
\langle D_+D_-\s(k_\perp^\pm)\rangle_N &= - \frac{h^*}{4\,\pi^2} \sum_{a = \pm} \left[ \frac{1}{(s_\perp^a)^2} + \frac{h^*g_2^*}{8\,\pi^2} \left( \frac{\log |s_\perp^a|}{(s_\perp^a)^2} + \frac{1}{2\,R|s_\perp^{-a}|} \log\left|\frac{s_\perp^a}{2\,R}\right| \right) \right] \ .
\end{aligned}
\end{equation*}
In the expansion of $L^+_-$ in $R$ we find a logarithmic divergence
\begin{equation} \label{log R}
\begin{aligned}
\de\langle D_+D_-\s(x)\rangle^{(1)}_N &\ni -\frac{(h^*)^2g_2^*}{32\,\pi^4}\log(R) \sum_{n\geq 0}\frac{R^{2\,n}}{x_\perp^{2(n + 1)}} \ .
\end{aligned}
\end{equation}
Note that this is not an IR divergence since $R$ is a distance scale. Still it should not be absorbed in the bare couplings on $D_\pm$. 

\begin{figure}
	\centering
	\includegraphics[width=.13\linewidth]{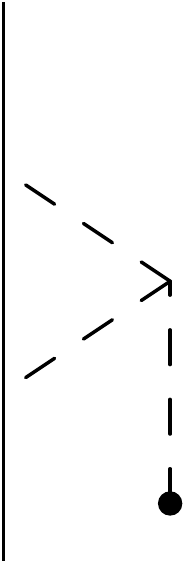}
	\caption{The single Feynman diagram at $\mco(g_2)$ in $\langle D_f\s(x)\rangle$.}
	\label{fig: diag 2}
\end{figure}

To avoid this logarithmic divergence we instead expand the integrands \eqref{Int Lab} of $L^a_b$ in $R$ before we integrate over $z_\perp$.
In this way we capture the logarithmic divergence in $R$ as a pole in $\e$
\begin{equation} \label{D+D- s}
\begin{aligned}
\langle D_+D_-\s(x)\rangle^{(1)}_N &= -h^2g_2A_d^3\pi^\frac{3\,p}{2} \frac{\G_{\De_\ph - \frac{p}{2}}^3}{\G_{\De_\ph}^3} \sum_{n\geq 0} a_nR^{2\,n}J^{d - p}_{\De_\ph - \frac{p}{2}, 2\,\De_\ph - p + n}(-x_\perp, 0) \ ,
\end{aligned}
\end{equation}
where $J^n_{a,b}(z, w^2)$ is the master integral \eqref{Master int 2}, and $a_n$ is the constant
\begin{equation}
\begin{aligned}
a_n &= \frac{(2\,\De_\ph - p)_n}{n!} + \frac{(4\,\De_\ph - 2\,p)_{2\,n}}{(2\,n)!} = \frac{2(n + 1)(2\,n^2 + 4\,n + 3)}{3} + \mco(\e) \ .
\end{aligned}
\end{equation}
We will now compute $\langle D_f\s(x)\rangle^{(1)}_N$ and see that it exactly equals \eqref{D+D- s}. For one-dimensional $R$ \eqref{one dim R}, it is given by
\begin{equation} \label{Df2}
\begin{aligned}
D_f = \exp\left( -2\,h \sum_{n \geq 0} \frac{R^{2\,n}}{(2\,n)!}\int_{\R^p}d^px\p_R^{2\,n}\hs(x_+) \right) \ .
\end{aligned}
\end{equation}
$\langle D_f\s(x)\rangle^{(1)}_N$ is found from the single Feynman diagram in Fig. \ref{fig: diag 2}
\begin{equation}
\begin{aligned}
\langle D_f\s(x)\rangle^{(1)}_N &= -\frac{(2\,h)^2g_2}{2} \int_{\R^d}d^dz \left.\langle \s(x)\s(z) \rangle\right|_{h = 0} \times \\
\eq \times \prod_{i = 1}^{2}\sum_{m_i\geq 0}\frac{R^{2\,m_i}}{(2\,m_i)!}\int_{\R^p}d^py_i \lim\limits_{R_i\rightarrow 0}\p_{R_i}^{2\,m_i}\left.\langle \s(z)\s(y_i \hx_\para +  R_i\hx_\perp^1) \rangle\right|_{h = 0} \ .
\end{aligned}
\end{equation}
Performing the integration over the parallel coordinates, and differentiating \eqref{b const}
gives us
\begin{equation} \label{c const}
\begin{aligned}
\langle D_f\s(x)\rangle^{(1)}_N &= -2\,h^2g_2A_d^3\pi^\frac{3\,p}{2} \frac{\G_{\De_\ph - \frac{p}{2}}^3}{\G_{\De_\ph}^3} \sum_{n\geq 0} c_nR^{2\,n} J^{d - p}_{\De_\ph - \frac{p}{2}, 2\,\De_\ph - p + n}(-x_\perp, 0) \ , \\
c_n &= \sum_{m = 0}^n b_mb_{n - m} = \frac{a_n}{2} \ .
\end{aligned}
\end{equation}
This is exactly the same as \eqref{D+D- s}
\begin{equation} \label{Fusion holds}
\begin{aligned}
\langle D_f\s(x)\rangle^{(1)}_N = \langle D_+D_-\s(x)\rangle^{(1)}_N \ .
\end{aligned}
\end{equation}
Thus the fusion \eqref{Df2} seems to hold even in the interacting theory. Note that using $D_f$, instead of $D_\pm$, simplified the Feynman diagram calculation as we did not need to calculate $L^+_-$ in \eqref{L + -} .

At this order $\langle D_f\s(x)\rangle^{(1)}_N$ has a single pole in $\e$ (neglecting constants of $\mco(\e^0)$)
\begin{equation} \label{D+D- s 2}
\begin{aligned}
\langle D_+D_-\s(x)\rangle^{(1)}_N &=  -\frac{h^2g_2}{16\,\pi^4x_\perp^2} \left( \frac{1}{\e} + \log(x_\perp^2) \right) + \\
\eq + \frac{h^2g_2}{96\,\pi^4} \sum_{n\geq 1} \frac{2\,n^2 + 4\,n + 3}{n}\frac{R^{2\,n}}{x_\perp^{2(n + 1)}} + \mco(\e) \ .
\end{aligned}
\end{equation}
The sum over $n$ was done in the thesis \cite{SoderbergRousu:2023ucv}.

\subsection{Order $g^2$}

We will now proceed to the next order in the bulk couplings, and see that fusion still holds. At this order, we will have $\mco(h\,g^2)$- and $\mco(h^3g^2)$-terms. Let us start with the former ones, which are only affected by one of the defects \eqref{Ds 2}
\begin{equation} \label{DDs g2h}
\begin{aligned}
\left.\langle D_+D_-\s \rangle^{(2)}_N\right|_{h\,g^2} &= -\frac{(N\,g_1^2 + g_2^2)h}{2} \left( A + A|_{R \rightarrow -R} \right) \ ,
\end{aligned}
\end{equation}
where the Fourier transform of $A$ is given by \eqref{ints A and B}. If we take the inverse we find
\begin{equation}
\begin{aligned}
A = \int_{\R^{d - p}}\frac{d^{d - p}k_\perp}{(2\,\pi)^{d - p}}e^{-i\,k_\perp(x_\perp - R)}\tilde{A} = \frac{\G_{2 - \frac{d}{2}} \G_{d - 5} }{ \G_{4 - \frac{d}{2}} \G_{d - 2} \G_{\frac{d - 2}{2}}^2 } \frac{\pi^{d + 1} A_d^4 }{|x_\perp - R|^{2(d - 5)}} \ .
\end{aligned}
\end{equation}
With this at hand we can expand the one-point function at \eqref{DDs g2h} in $R$ 
\begin{equation} \label{DDs g2h 2}
\begin{aligned}
\left.\langle D_+D_-\s \rangle^{(2)}_N\right|_{h\,g^2} &= -\frac{32\,h(N\,g_1^2 + g_2^2)\pi^{d + 1}A_d^4}{(d - 4)^2(d^2 - 9\,d + 18) \G_{2\,d - 9} \G_{\frac{d}{2} - 2}^2 } \sum_{n\geq 0} \frac{\G_{\frac{d + 5 - 5}{2}}}{(2\,n)!} \frac{R^{2\,n}}{x_\perp^{2(d - 5 + n)}} \ .
\end{aligned}
\end{equation}
We wish to point out that if we were to expand in $R$ before doing the integrals \eqref{ints A and B 0} over $z_\perp^i$ in $A$ we find divergences at $\mco(R^{2\,n})$ that go as $H_{-n}$ (the harmonic number) which we cannot regulate using dimensional regularization. 

\begin{figure}
	\begin{subfigure}{.5\textwidth}
		\centering
		\includegraphics[width=.5\linewidth]{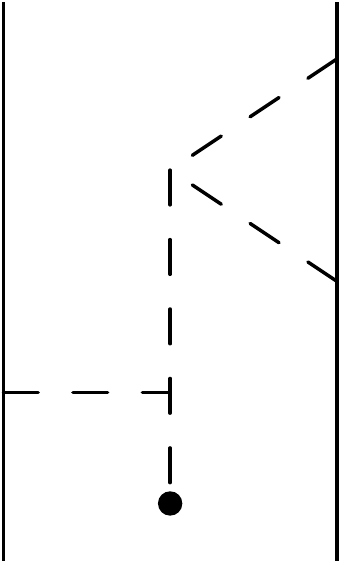}
	\end{subfigure}%
	\begin{subfigure}{.5\textwidth}
		\centering
		\includegraphics[width=.5\linewidth]{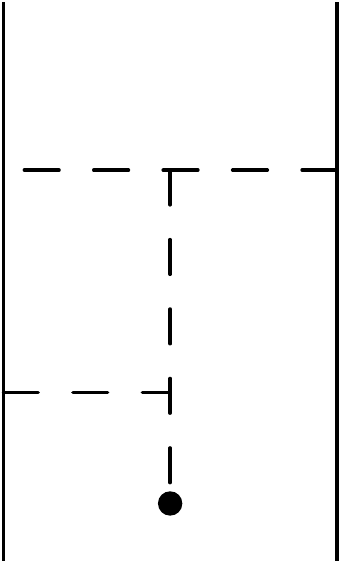}
	\end{subfigure}
	\caption{The two Feynman diagrams that contribute to the one-point function of $\s$ (in the presence of the two defects $D_\pm$) at $\mco(g^2)$.}
	\label{fig: g2Fusion}
\end{figure}

The corresponding part of $\langle D_f\s\rangle$ can be found from the first diagram in Fig. \ref{fig: g2} (neglecting the defect not connected to the vertex)
\begin{equation*}
\begin{aligned}
\left.\langle D_f\s(x)\rangle^{(2)}_N\right|_{h\,g^2} &= (-2\,h)\frac{N\,g_1^2 + g_2^2}{2}
\int_{\R^d}d^dz\int_{\R^d}d^dw \left.\langle \s(x)\s(z) \rangle\right|_{h = 0}\left.\langle \s(z)\s(w) \rangle^2\right|_{h = 0} \times \\
\eq\times \sum_{n\geq 0}\frac{R^{2\,n}}{(2\,n)!}\int_{\R^p}d^py \lim\limits_{R'\rightarrow 0}\p_{R'}^{2\,n}\left.\langle \s(w)\s(y \hx_\para +  R'\hx_\perp^1) \rangle\right|_{h = 0} \ .
\end{aligned}
\end{equation*}
We can integrate over the parallel coordinates using the master integral \eqref{Master int 1}, and over the normal coordinates using \eqref{Master int 2}. After this we can differentiate w.r.t. $R'$ using \eqref{b const}
\begin{equation} \label{Dfs hg2}
\begin{aligned}
\left.\langle D_f\s(x)\rangle^{(2)}_N\right|_{h\,g^2} &= -h(N\,g_1^2 + g_2^2)\pi^{d + \frac{p}{2}}A_d^4 \frac{ \G_{\frac{d}{2} - 2\,\De_\ph} \G_{4\,\De_\ph - d - \frac{p}{2}} \G_{\frac{d}{2} - \De_\ph}^2 }{ \G_{\frac{3\,d}{2} - 4\,\De_\ph} \G_{2\,\De_\ph} \G_{\De_\ph}^2 } \times \\
\eq\times \sum_{n\geq 0}\frac{(8\,\De_\ph - 2\,d - p)_n}{(2\,n)!} \frac{R^{2\,n}}{|x_\perp|^{8\,\De_\ph - 2\,d - p + 2\,n}} \\
&= \left.\langle D_+D_-\s(x)\rangle^{(2)}_N\right|_{h\,g^2} \ ,
\end{aligned}
\end{equation}
which is in exact agreement with \eqref{DDs g2h 2} (since $\De_\ph = \frac{d - 2}{2}$, $p = 2$ in the Feynman diagrams). It has a single pole in $\e$ (neglecting constants at $\mco(\e^0)$)
\begin{equation} \label{exp 2}
\begin{aligned}
\left.\langle D_f\s(x)\rangle^{(2)}_N\right|_{h\,g^2} &= \frac{( N\,g_1^2 + g_2^2 )h}{768\, \pi^5} \sum_{n\geq 0} (2\,n + 1) \frac{R^{2\,n}}{|x_\perp|^{2(n + 1)}} \left( \frac{1}{\e} + \log(x_\perp^2) + \mco(\e) \right) \ .
\end{aligned}
\end{equation}
The other part of $\langle D_+D_-\s(x)\rangle^{(2)}_N$ are those of $\mco(h^3g^2)$, which contain the $B$-part (here we denote $B = C_{+, +, +}$) of the the one-point function \eqref{Ds 2} in the presence of only one defect as well as the two connecting diagrams in Fig. \ref{fig: g2Fusion}
\begin{equation} \label{DDs h3g2}
\begin{aligned}
\left.\langle D_+D_-\s \rangle^{(2)}_N\right|_{h^3g^2} &= (3!)^2(-h)^3\left( -\frac{g_2}{3!} \right)^2 \bigg( \frac{C_{+,+,+} + C_{-,-,-} + C_{+,-,-} + C_{-,+,+}}{2!} + \\
\eq + C_{+,+,-} + C_{-,-,+} \bigg) \ .
\end{aligned}
\end{equation}
This is written in terms of the integral
\begin{equation}
\begin{aligned}
C_{\al,\be,\g} &= \int_{\R^d}d^dz\int_{\R^d}d^dw \left.\langle \s(x)\s(z) \rangle\right|_{h = 0} \left.\langle \s(z)\s(w) \rangle\right|_{h = 0} K_\al(z) K_\be(z) K_\g(z) \\
&= \left( \pi^\frac{p}{2}A_d\frac{\G_{\De_\ph - \frac{p}{2}}}{\G_{\De_\ph}} \right)^5\int_{\R^{d - p}}d^{d - p}z_\perp \int_{\R^{d - p}}d^{d - p}w_\perp \times \\
\eq \times \frac{1}{(|x_\perp - z_\perp|\,|z_\perp - w_\perp|\,|x_\perp - \al R|\,|x_\perp - \be R|\,|x_\perp - \g R|)^{2\,\De_\ph - p}} \ .
\end{aligned}
\end{equation}
With this at hand we can expand \eqref{DDs h3g2} in $R$ and integrate over the normal coordinates
\begin{equation} \label{DDs h3g2 v2}
\begin{aligned}
\left.\langle D_+D_-\s \rangle^{(2)}_N\right|_{h^3g^2} &= -4\,\pi^{\frac{d}{2} 2\,p} A_d^5 h^3g_2^2 \frac{ \G_{ \frac{d + p}{2} - 2\,\De_\ph } \G_{ \frac{d}{2} - \De_\ph } \G_{ 3\,\De_\ph -\frac{d}{2} - p } \G_{\De_\ph -\frac{p}{2}}^4 }{ \G_{ d + \frac{p}{2} - 3\,\De_\ph } \G_{ 2\,\De_\ph - p } \G_{ \De_\ph }^5 } \times \\
\eq\times J^{d - p}_{4\,\De_\ph - \frac{d + 3\,p}{2}, \De_\ph - \frac{p}{2}}(x_\perp) + \mco(R^2)  \ .
\end{aligned}
\end{equation}
On the other hand, for $D_f$ we have
\begin{equation*}
\begin{aligned}
\left.\langle D_f\s\rangle^{(2)}_N\right|_{h^3g^2} &= \frac{(3!)^2}{2!}(-2\,h)^3 \left( -\frac{g_2}{3!} \right)^2 \int_{\R^d}d^dz \int_{\R^d}d^dw \left.\langle \s(x)\s(z) \rangle\right|_{h = 0} \left.\langle \s(z)\s(w) \rangle\right|_{h = 0} \times \\
\eq\times \prod_{i = 1}^3 \sum_{m_i \geq 0} \frac{R^{2\,m_i}}{(2\,m_i)!} \int_{\R^p}d^py_i \lim\limits_{R_1 \rightarrow 0} \left.\langle \s(z)\s(y_i\hx_\para + R_1\hx_\perp^1) \rangle\right|_{h = 0} \times \\
\eq\times \prod_{j = 2}^2 \lim\limits_{R_j \rightarrow 0} \left.\langle \s(w)\s(y_j\hx_\para + R_j\hx_\perp^1) \rangle\right|_{h = 0} \ .
\end{aligned}
\end{equation*}
To solve this integral we perform the following steps:
\begin{enumerate}
	\item Integrate over the parallel coordinates.
	\item Differentiate w.r.t. $R_i$.
	\item Integrate over $w_\perp$.
	\item Integrate over $z_\perp$.
\end{enumerate}
Doing this gives us perfect agreement with \eqref{DDs h3g2 v2} (order by order in $R$)
\begin{equation} \label{Dfs h3g2}
\begin{aligned}
\left.\langle D_f\s\rangle^{(2)}_N\right|_{h^3g^2} &= -\frac{128\,\pi^{d + 3}A_d^5h^3g_2^2}{\G_{\frac{d - 2}{2}}^2\G_{d - 3}^3} \sum_{n\geq 0}c_n\frac{R^{2\,n}}{|x_\perp|^{3\,d - 16 + 2\,n}} = \left.\langle D_+D_-\s\rangle^{(2)}_N\right|_{h^3g^2} \ .
\end{aligned}
\end{equation}
The constant $c_n$ is given by a finite sum
\begin{equation*}
\begin{aligned}
c_n &= \sum_{m = 0}^n\sum_{m' = 0}^m a_{n - m, m - m', m'} \\
&= \frac{4}{(3\,d + 2\,n - 16)(d + n - 6)\G_{d - 4}^2} \sum_{m = 0}^n\frac{\G_{d + 2(n - m) - 4} \G_{d + 2\,m - 4} }{ (d + m - 5)(d + 2\,m - 6)\G_{2(n + m) + 1} \G_{2\,m + 1} } \times \\
\eq\times {}_4F_3\left( \frac{d - 2}{2}, \frac{d - 3}{2}, \frac{1}{2} - m, -m; \frac{1}{2}, \frac{5 - d}{2} - m, 3 - \frac{d}{2} - m; 1  \right) \ .
\end{aligned}
\end{equation*}
Here $a_{m_1, m_2, m_3}$ is
\begin{equation}
\begin{aligned}
a_{m_1, m_2, m_3} &= b_{m_1}b_{m_2}b_{m_3} \frac{ \G_{d + p - 4\,\De_\ph - m_1 - m_2 - m_3} \G_{3\,\De_\ph - \frac{d}{2} - p + m_2 + m_3}  }{ \G_{\frac{3\,d}{2} + p - 5\,\De_\ph - m_1 - m_2 - m_3} \G_{2\,\De_\ph - p + m_2 + m_3}  } \times \\
\eq\times \frac{  \G_{5\,\De_\ph - d - \frac{3\,p}{2} + m_1 + m_2 + m_3} \G_{\frac{d + p}{2} - 2\,\De_\ph - m_2 - m_3} }{  \G_{4\,\De_\ph - \frac{d + 3\,p}{2} + m_1 + m_2 + m_3} \G_{d + \frac{p}{2} - 3\,\De_\ph - m_2 - m_3} } \ .
\end{aligned}
\end{equation}
$\left.\langle D_f\s\rangle^{(2)}_N\right|_{h^3g^2}$ has the $\e$-expansion (we do not care about constants at $\mco(\e^0)$)
\begin{equation} \label{exp 3}
\begin{aligned}
\langle D_f\s\rangle^{(2)}_N &= -\frac{h^3g_2^2}{128\,\pi^6x_\perp^2} \left( \frac{1}{\e^2} + 3\frac{\mathcal{A} + \log(x_\perp^2)}{2\,\e} + \frac{9}{2}\left( \frac{\log(x_\perp^2)}{4} \right)^2 + \frac{9\,\mathcal{A}\,\log(x_\perp^2)}{4} + \rig \\
\eq\lef + \sum_{n\geq 0} \frac{2\,n + 1}{n(n + 1)} \left( \frac{1}{\e} + 3\frac{\log(x_\perp^2)}{2} \right) + \mco(\e) \right) \ , \\
\mathcal{A} &= \log\left( \pi\,e^{\g_E + \frac{5}{3}} \right) \ .
\end{aligned}
\end{equation}
To summarize this Section, we found (by studying terms of the same order in $h$) in (\ref{Dfs hg2}, \ref{Dfs h3g2}) that fusion holds at $\mco(g^2)$
\begin{equation}
\begin{aligned}
\langle D_f\s\rangle^{(2)}_N &= \langle D_+D_-\s\rangle^{(2)}_N \ .
\end{aligned}
\end{equation}

\subsection{Renormalization}

In this Section we will renormalize the one-point function (\ref{c const}, \ref{Dfs hg2}, \ref{Dfs h3g2}) of $\s$ in the presence of the fused defect, $D_f$. To do this we treat each order in $R$ on $D_f$ \eqref{Df2} as independent couplings
\begin{equation} \label{Bare Df}
\begin{aligned}
D_f = \exp\left( -2 \sum_{n \geq 0} h_n\frac{R^{2\,n}}{(2\,n)!}\int_{\R^p}d^px\p_R^{2\,n}\hs(x_+) \right) \ ,
\end{aligned}
\end{equation}
where $h_n$ is the set of bare couplings on $D_f$. For $n \geq 1$ these are dimensionfull ($R^{2\,n}$), and thus we expect these to not have any non-trivial f.p.'s. In this Section we will see that this is indeed the case. This will in turn mean that after we have fused the two (conformal) defects, $D_\pm$, we can turn on the bulk-interactions and flow to a conformal f.p. in the RG where $D_f$ is also conformal. 

The one-point function near $D_f$ is given by (\ref{c const}, \ref{Dfs hg2}, \ref{Dfs h3g2}). To avoid $\frac{\g_E}{\e}$-terms (of $\mco(h^3g_2^2)$) after renormalization we have to factor out the free theory contribution. Let us write $\langle D_f\s\rangle_N$ in the following way
\begin{equation}
\begin{aligned}
\langle D_f\s\rangle_N &= D_0 + \sum_{n\geq 0}R^{2\,n}D_n + \mco(g^3) \ .
\end{aligned}
\end{equation}
The $\mco(R^0)$-terms are given by
\begin{equation}
\begin{aligned}
D_0 &= -\frac{h_0\G_{1 - \frac{\e}{2}}}{2\,\pi^{2 - \frac{\e}{2}}|x_\perp|^{2 - \e}} \left( 1 + \frac{g_2h_0\G_{-\frac{\e}{2}}|x_\perp|^\e}{16\,\pi^{2 - \frac{\e}{2}}(\e - 1)} + \frac{(N\,g_1^2 + g_2^2)\G_{-\frac{\e}{2}}|x_\perp|^\e}{256\,\pi^{3 - \frac{\e}{2}}(\e^2 - 4\,\e + 3)} + \rig \\
\eq\lef + \frac{g_2^2h_0^2\G_{-\frac{\e}{2}}^2|x_\perp|^{2\,\e}}{128(\e^2 - 5\,\e + 2)} \right) \ ,
\end{aligned}
\end{equation}
and the $\mco(R^{2\,n})$-terms are
\begin{equation*}
\begin{aligned}
D_n &= -\frac{h_n\G_{1 - \frac{\e}{2}}(2 - \e)_n}{2\,\pi^{2 - \frac{\e}{2}}(2\,n)!|x_\perp|^{2(n + 1) - \e}} \bigg[ \left. 1 + \rig \\
\eq\lef - \frac{4^\e\pi\G_{2(n + 2 - \e)} + 2^{3 + 2\,n} \G_{n +\frac{1}{2}} \G_{\frac{5}{2} - \e} \G_{n + 2 - \e} }{ \G_{\frac{5}{2} - \e} \G_{n + 2 - \e} }\frac{ g_2h_n \G_{1 - \frac{\e}{2}} |x_\perp|^\e }{128\,\pi^{\frac{5 - \e}{2}}(2\,n - \e)(n + 1- \e) } + \rig \\
\eq\lef - \frac{\G_{2(n + 1 - \e)}\G_{-\frac{\e}{2}}}{\G_{\frac{3}{2} - \e}\G_{n + 2 - \e}} \frac{(N\,g_1^2 + g_2^2)|x_\perp|^\e}{2^{9 - 2\,\e}\pi^{5 - \frac{\e}{2}}(\e - 3) } + \rig \\
\eq\lef + \frac{\G_{2\,n}\G_{1 - \frac{\e}{2}}^2}{\G_{2 - \e}\G_{n + 2 - \e}} \frac{g_2^2h_n^2n|x_\perp|^{2\,\e}}{16\,\pi^{4 - \e}(3\,\e + 2\,n + 2)(\e - n)} \sum_{m = 0}^n \frac{\G_{2(n - m + 1) - \e}\G_{2(m + 1) - \e}}{ \G_{2(n - m) + 1} \G_{2\,m + 1} } \times \rig \\
\eq\lef\times \frac{ 1 }{ (\e - 2\,m) (\e - m - 1) } {}_4F_3 \left( 1 - \frac{\e}{2}, \frac{3 - \e}{2}, \frac{1}{2} - m, -m; \frac{1}{2}, \frac{\e - 1}{2} - m, \frac{\e}{2} - m ; 1 \right) \right. \bigg] \ .
\end{aligned}
\end{equation*}
Here $(x)_n$ is the Pochhammer symbol. $\langle D_f\s\rangle_N$ admit the $\e$-expansions at (\ref{D+D- s 2}, \ref{exp 2}, \ref{exp 3}). To renormalize $D_0$, $D_n$ we need the bare bulk couplings, $g_i$, at \eqref{bare def} and the $Z$-factor (\ref{Z-factor 1}, \ref{Z-factor 2}) for $\s$. For the same reason as in Sec. \ref{Sec: fp} (see the discussion above \eqref{renorm s}) we divide each bare defect coupling in the one-point function of the renormalized field $\tilde{\s}$ with $\sqrt{Z}$: $h_n \rightarrow \frac{h_n}{\sqrt{Z}}$. Finally, we make the following ansatz for the bare defect couplings
\begin{equation} \label{fused bare}
\begin{aligned}
h_n = \m^{\frac{\e}{2}}\tilde{h}_n\left( 1 + a_n\frac{\tilde{h}_n\tilde{g}_2}{\e} + b_n\frac{N\,\tilde{g}_1^2 + \tilde{g}_2^2}{\e} + c_n^2\frac{\tilde{h}_n^2\tilde{g}_2^2}{\e^2}  + c_n^1\frac{\tilde{h}_n^2\tilde{g}_2^2}{\e} + \mco(g^3) \right) \ .
\end{aligned}
\end{equation}
By cancelling the poles in $\e$ we are able to fix the constants $a_n$, $b_n$, $c_n^2$, $c_n^1$ (note that some of these are zero)
\begin{equation*}
\begin{aligned}
h_0 &= \m^{\frac{\e}{2}}\tilde{h}_0\left( 1 - \frac{\tilde{h}_0\tilde{g}_2}{8\,\pi^2\e} + \frac{\tilde{h}_n^2\tilde{g}_2^2}{64\,\pi^4\e^2} - \frac{\tilde{h}_n^2\tilde{g}_2^2}{128\,\pi^4\e} + \mco(g^3) \right) \ , \\
h_{n\geq 1} &= \m^{\frac{\e}{2}}\tilde{h}_n\left( 1 + \left( \frac{ 2^{2\,n - 7} (\frac{3}{2})_n }{3(n + 1)\pi^3} - \frac{1}{384\,\pi^3} \right) \frac{N\,\tilde{g}_1^2 + \tilde{g}_2^2}{\e} + \frac{2^{2\,n - 5}\G_{n + \frac{3}{2}}}{\pi^\frac{9}{2} n(n + 1)^2 }\frac{\tilde{h}_n^2\tilde{g}_2^2}{\e} + \mco(g^3) \right) \ ,
\end{aligned}
\end{equation*}
where there are a couple of consistency checks to be made. Firstly, there is no $N\,g_1^2$-term in $h_0$ (it was exactly cancelled by the $Z$-factor \eqref{Z-factor 2} of $\s$). This is expected as otherwise it would affect its f.p. found at the $\mco(g)$. Secondly, $c_0^2 = a_0^2$ which is required to cancel a pole of $\e$ in the $\be$-function of $h_0$. Thirdly, the renormalized one-point function, $\langle\tilde{\s}\rangle$, only contain dimensionless logarithms, $\log(x_\perp^2\m^2)$, as expected.

The $\be$-functions for the fused defect couplings are 
\begin{equation} \label{Df beta d = 6}
\begin{aligned}
\be_0 &= -\frac{\tilde{h}_0}{\e} - \frac{\tilde{h}_0^2\tilde{g}_2}{8\,\pi^2} - \frac{\tilde{h}_0^3\tilde{g}_2^2}{64\,\pi^4} \ , \\
\be_{n\geq 1} &= -\frac{\tilde{h}_n}{\e} + \left( \frac{ 2^{2\,n - 7} (\frac{3}{2})_n }{3(n + 1)\pi^3} - \frac{1}{384\,\pi^3} \right) (N\,\tilde{g}_1^2 + \tilde{g}_2^2)\tilde{h}_n + \frac{4^{n - 2}\G_{n + \frac{3}{2}}}{\pi^\frac{9}{2} n(n + 1)^2 }\tilde{h}_n^3\tilde{g}_2^2 \ .
\end{aligned}
\end{equation}
From which we can find their f.p.'s. Aside from the trivial f.p.'s
\begin{equation} \label{triv fp}
\begin{aligned}
h_0^* &= 0 \ , \quad h_n^* = 0 \ ,
\end{aligned}
\end{equation}
there are also the non-trivial ones
\begin{equation}
\begin{aligned}
h_0^* &= -4\,\pi^2 \left( \frac{1}{\tilde{g}_2} \mp \sqrt{\frac{1 - 2\,\e}{\tilde{g}_2^2}} \right) \ , \\
h_{n\geq 1}^* &= \pm i\frac{\pi^{\frac{3}{4}}}{2^{n + 3} 3^\frac{3}{2} } \sqrt{ n(n + 1) \frac{2^{2\,n + 1}(\frac{3}{2})_n - 3(n + 1) }{\G_{n + \frac{3}{2}}} } \sqrt{N} + \mco\left(\frac{1}{N^2}\right) \ .
\end{aligned}
\end{equation}
The f.p. for $h_{n\geq 1}$ is not perturbative, and thus only its trivial f.p. at \eqref{triv fp} is valid. The dimensionless coupling, $h_0$, on the other hand has a non-trivial and attractive f.p. given by half \eqref{Def fp} of that on $D_\pm$
\begin{equation} \label{Df fp d = 6}
\begin{aligned}
\left(h_0^*, h_{n\geq 1}^*\right) &= \left( \frac{h^*}{2}, 0 \right) \ .
\end{aligned}
\end{equation}
Specifying to this non-trivial f.p., we find the renormalized one-point function
\begin{equation}
\begin{aligned}
\langle\tilde{\s}(x)\rangle &= -\frac{h_0^*\G_{1 - \frac{\e}{2}}}{2\,\pi^{2 - \frac{\e}{2}}|x_\perp|^{2 - \e}} \left( 1 + \frac{g_2^*h_0^*}{(4\,\pi^2)^2}\log(x_\perp^2\m^2) - \frac{N(g_1^*)^2 + (g_2^*)^2}{768\,\pi^3} \log(x_\perp^2\m^2) + \rig \\
\eq\lef + \frac{(g_2^*)^2(h_0^*)^2}{(4\,\pi)^4} \left( \log( x_\perp^2\m^2) + \mathcal{C} \right)^2 + \mco(g^3) \right) \ , \\
\mathcal{C} &= \log\left( \frac{\sqrt{\pi}}{e^{\frac{\g_E - 3}{2}}} \right) \ .
\end{aligned}
\end{equation}
From this expression we find the correct bulk anomalous dimension of $\s$ \cite{Fei:2014yja}
\begin{equation}
\begin{aligned}
\De_\s = \frac{d - 2}{2} + (N(g_1^*)^2 + (g_2^*)^2)\g_\s + \mco(g^3) \ , \quad \g_s = \frac{1}{768\,\pi^3} \ .
\end{aligned}
\end{equation}


\subsection{Line defects near four dimensions}

We will end this Section by showing that fusion also seems to hold at a quantum level in $d = 4 - \e$. This calculation is similar to that in Sec. \ref{Sec: one-pt fcn}. Near four dimensions we can consider a quartic bulk-interaction invariant under $O(N)$
\begin{equation}
\begin{aligned}
S = \int_{\R^d}d^dx \left( \frac{(\p_\m\ph^i)^2}{2} + \frac{\la}{8}\ph^4 \right) \ , 
\end{aligned}
\end{equation}
where $\ph^4 \equiv [(\ph^i)^2]^2$ and $i \in \{1, .., N\}$. In addition we consider two $p = 1$ dimensional line defects on the form \eqref{Defects} (without the $\s$-term). The bulk interactions has the non-trivial WF f.p.
\begin{equation}
\begin{aligned}
\la^* = \frac{(4\,\pi)^2\e}{N + 8} + \mco(\e^2) \ ,
\end{aligned}
\end{equation}
and on the defects we have the following attractive f.p.'s \cite{Cuomo:2021kfm}
\begin{equation} \label{4d def fp}
\begin{aligned}
h^*_\pm = \pm\sqrt{N + 8} \pm \frac{4\,N^2 + 45\,N + 170}{4(N + 8)^\frac{3}{2}}\e + \mco(\e^2) \ ,
\end{aligned}
\end{equation}
which is of finite size.

Fusing the defects \eqref{Defects} with a multivariate Taylor expansion yields
\begin{equation}
\begin{aligned}
D_f = \exp\left( -h \prod_{i = 1}^{d - p} \sum_{n_i \geq 0} \frac{\de^{ji_+} + (-1)^n\de^{ji_-}}{n_i!}R_i^{n_i}\int_{\R^p}d^px\p_i^{n_i}\hp^j(x_+) \right) \ .
\end{aligned}
\end{equation}
This reduces to the form \eqref{Df} when $i_\pm = N$ (under the exchange $\s \rightarrow \ph^N$). 

\begin{figure}
	\begin{subfigure}{.5\textwidth}
		\centering
		\includegraphics[width=.5\linewidth]{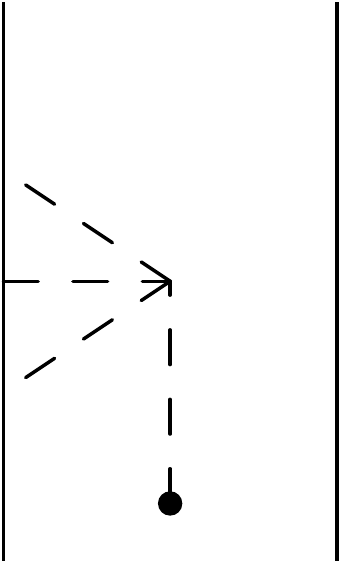}
	\end{subfigure}%
	\begin{subfigure}{.5\textwidth}
		\centering
		\includegraphics[width=.5\linewidth]{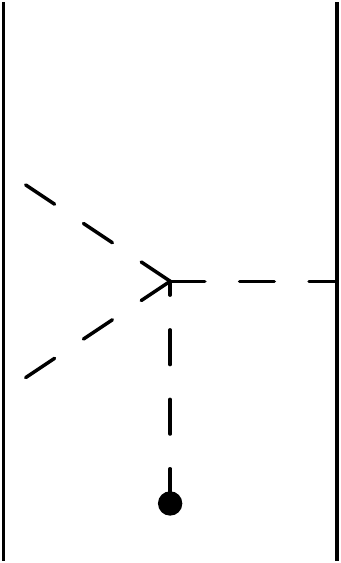}
	\end{subfigure}
	\caption{The diagrams that contribute to the one-point function of $\ph^i$ in $d = 4 - \e$.}
	\label{fig: 4d diag}
\end{figure}

At $\mco(\la)$ we find $\langle D_+D_-\ph^i(x)\rangle^{(1)}_N$ from the Feynman diagrams in Fig. \ref{fig: 4d diag}
\begin{equation*}
\begin{aligned}
\langle D_+D_-\ph^i(x)\rangle^{(1)}_N &= (-h)^3\left(-\frac{\la}{8}\right) \sum_{a = \pm} \left( \frac{4!}{3!}\de^{i\,i_a}\tilde{L}^a_a  + \frac{8!}{2}(2\,\de^{i_+i_-}\de^{i\,i_{+a}} + \de^{i\,i_{-a}})\tilde{L}^{+a}_{-a} \right) \ .
\end{aligned}
\end{equation*}
This is expressed in terms of the following integral
\begin{equation}
\begin{aligned}
\tilde{L}^a_b = \int_{\R^d}d^dz\left.\langle\ph(x)\ph(z)\rangle\right|_{h = 0}K_a(z)^2K_b(z) \ .
\end{aligned}
\end{equation}
We will now perform the following steps:
\begin{enumerate}
	\item Integrate over the parallel coordinate, $z_\para \in \R$, in $\tilde{L}^a_b$.
	\item Expand in $R$.
	\item Integrate over $z_\perp \in \R^{d - 1}$.
\end{enumerate}
Doing this yields
\begin{equation} \label{4d DD}
\begin{aligned}
\tilde{L}^a_b &= \pi^{2\,p}A_d^4\frac{\G_{\De_\ph - \frac{p}{2}}^4}{\G_{\De_\ph}^4} \left[ J^{d - p}_{\De_\ph - \frac{p}{2}, 3\,\De_\ph - \frac{3\,p}{2}}(-x_\perp, 0) + \rig \\
\eq \lef + (2\,a + b)(p - 2\,\De_\ph)R\,J^{d - p}_{\De_\ph - \frac{p}{2}, 3\,\De_\ph - \frac{3\,p}{2} + 1}(-x_\perp, 0) + \mco(R^2) \right] \ .
\end{aligned}
\end{equation}
On the other hand, for $D_f$ we have
\begin{equation} \label{4d Df}
\begin{aligned} 
\langle D_f\ph^i(x)\rangle^{(1)}_N &= \frac{4!}{3!}(-h)^3\left(-\frac{\la}{8}\right) \int_{\R^d}d^dz\left.\langle\ph(x)\ph(z)\rangle\right|_{h = 0} \prod_{i = 1}^3 \int_{\R^p}d^pw_i \times \\
\eq\times \sum_{n_i\geq 0} \s_{n_1, n_2, n_3} \frac{R^{2\,n_i}}{(2\,n_i)!}\lim\limits_{R_i \rightarrow 0}\p_i^{2\,n_i}\left.\langle\ph(z)\ph(w_i)\rangle\right|_{h = 0} \\
&= \frac{4\,\pi^{2\,p}A_d^4h^3\la\,\G_{\De_\ph - \frac{p}{2}}^4\de^{i\,N}}{\G_{\De_\ph}^4} \sum_{n\geq 0}d_nR^{2\,n}J^{d - p}_{\De_\ph - \frac{p}{2}, 3\,\De_\ph - \frac{3\,p}{2} + n}(-x_\perp, 0) \ , 
\end{aligned}
\end{equation}
with the constant
\begin{equation}
\begin{aligned}
d_n &= \sum_{m = 0}^n\sum_{m' = 0}^mb_{\frac{n - m}{2}}b_{\frac{m - m'}{2}}b_{\frac{m'}{2}}\s_{n - m, m - m', m'} \\
&= 154\frac{\de^{ii_+} + (-1)^n\de^{ii_-} }{ \G_{2\,\De_\ph - p}^2 } \sum_{\substack{m\geq 0 \\ \text{even } m}} \frac{1}{(n - 2\,m)!} \left( \frac{ \G_{ 2\,\De_\ph - p + m } \G_{ 2\,\De_\ph - p + n - 2\,m } }{m!} + \rig \\
\eq\lef - \frac{\pi\,\csc\left[ \pi(p - 2\,\De_\ph) \right]}{(2\,m)!} {}_2F_1 \left( -2\,m, 2\,\De_\ph - p ; p - 2\,\De_\ph - 2\,m + 1 ; 1 \right) \right) \ .
\end{aligned}
\end{equation}
Here $b_m$ is the constant in \eqref{b const} and $\s_{n_1, n_2, n_3}$ is a factor from applying Wick's theorem to the integrand
\begin{equation}
\begin{aligned} 
\s_{n_1, n_2, n_3} = \left( 1 + (-1)^{n_2 + n_3} + \left[ (-1)^{n_2} + (-1)^{n_3} \right] \de^{+-} \right)\left( \de^{i+} + (-1)^{n_1}\de^{i-} \right) \ .
\end{aligned}
\end{equation}
$\langle D_f\ph^i(x)\rangle^{(1)}_N$ at \eqref{4d Df} is in perfect agreement with $\langle D_+D_-\ph^i(x)\rangle^{(1)}_N$ at \eqref{4d DD} (seen order by order in $R$). This suggests that the fusion \eqref{Df2} is valid in $d = 4 - \e$ as well
\begin{equation}
\begin{aligned} 
\langle D_f\ph^i(x)\rangle^{(1)}_N &= \langle D_+D_-\ph^i(x)\rangle^{(1)}_N \ .
\end{aligned}
\end{equation}
If we expand $\langle D_f\ph^i(x)\rangle^{(1)}_N$ in $\e$ we find (neglecting constants of $\mco(\e^0)$)
\begin{equation} 
\begin{aligned}
\langle D_+D_-\ph^i(x)\rangle^{(1)}_N &= \frac{h^3\la\,\de^{i\,N}}{32\,\pi^3|x_\perp|} \left( \frac{1}{\e} + \frac{3}{2}\log(x_\perp^2)  + \frac{1}{2} \sum_{n \geq 0} \frac{(n + 1)(n + 2)}{n(2\,n + 1)}\frac{R^{2\,n}}{x_\perp^{2\,n}} \right) \ . 
\end{aligned}
\end{equation}
Writing $D_f$ as \eqref{Bare Df}, we find that $\langle D_f\ph^i(x)\rangle^{(1)}_N$ is renormalized by the following bare couplings constants\footnote{At $\mco(\la)$ the bulk-field, $\ph^i$, receives no anomalous dimension and thus we do not have to bother with a $Z$-factor \cite{WILSON1974119}.}
\begin{equation}
\begin{aligned}
h_0 = \m^\frac{\e}{2} \tilde{h}_0 \left( 1 + \frac{\tilde{h}_0^2\tilde{\la}}{8\,\pi^2\,\e} + \mco(\tilde{\la}^2) \right) \ , \quad  h_n = \m^\frac{\e}{2} \tilde{h}_n + \mco(\tilde{\la}^2) \ .
\end{aligned}
\end{equation}
which gives us the $\be$-functions
\begin{equation} \label{Df beta d = 4}
\begin{aligned}
\be_0 = -\frac{\e}{2}\tilde{h}_0 + \frac{\tilde{h}_0^3\tilde{\la}}{4\,\pi^2} + \mco(\tilde{\la}^2) \ , \quad \be_n = -\frac{\e}{2}\tilde{h}_n + \mco(\tilde{\la}^2) \ .
\end{aligned}
\end{equation}
This $\be$-function has the non-trivial f.p.
\begin{equation} \label{Df fp d = 4}
\begin{aligned}
h_0^* = \pm\pi\sqrt{\frac{2\,\e}{\la*}} = \pm\sqrt{\frac{N + 8}{8}} + \mco(\e) \ , \quad h_n^* = 0 \ .
\end{aligned}
\end{equation}
Note that this differs from \eqref{4d def fp} by a factor of $\sqrt{8}$ at $\mco(\e^0)$. The renormalized one-point function is
\begin{equation*}
\begin{aligned}
\langle D_f\ph^i(x)\rangle^{(1)}_N &= \frac{3(h_f^*)^3\la^*\de^{i\,N}}{32\,\pi^3|x_\perp|} \log|x_\perp| + \mco(\e^2) \ ,
\end{aligned}
\end{equation*}
where
\begin{equation}
\begin{aligned}
(h_f^*)^3\la^* = \pm\sqrt{\frac{N + 8}{2}}\pi\,\e + \mco(\e^2) \ .
\end{aligned}
\end{equation}

\section{Conclusion}

In this paper we have fused two scalar Wilson defects \eqref{Scalar Wilson loop} in $d = 4 - \e$ and $d = 6 - \e$, and presented results which indicate that this fusion \eqref{fusion} also holds in the interacting theory. In particular, we showed that bulk one-point functions stay invariant (before renormalization). This is an expected result since the path integral is the same before and after fusion. 

From our results we see the power of fusion. Firstly it gives rise to an infinite tower of interactions \eqref{fusion}. However, as we have shown in this paper, the dimensionfull couplings does not have non-trivial f.p.'s and can thus be tuned to zero in conformal field theories. Assuming this to start with would greatly simplify the calculation of Feynman diagrams.

We found that the coupling constants on the fused defect, $D_f$, also takes into account divergences in the fusion-limit of the two defects, which might gives rise to different $\be$-functions (\ref{Df beta d = 6}, \ref{Df beta d = 4}). E.g. in $d = 4 - \e$ the fused defect f.p. \eqref{Df fp d = 4} differ by a factor of $\sqrt{8}$, while in $d = 6 - \e$ it stays the same \eqref{Df fp d = 6} (remember that there is a factor of $2$ in front of the fused defect couplings \eqref{fusion}). We believe the underlying reason to this can be seen from the OPE, wherein $d = 4 - \e$ the external field is not exchanged in its own OPE while this is the case in $d = 6 - \e$
\begin{equation}
\begin{aligned}
\ph \times \ph &\sim \1 + \ph^2 + ... \ , &\quad &\text{if } d = 4 - \e \ , \\
\s \times \s &\sim \1 + \s + ... \ , &\quad &\text{if } d = 6 - \e \ .
\end{aligned}
\end{equation}
Of course, a more detailed study on the relation between the (bulk-bulk or defect-defect) OPE's and fusion is required. 

However, both of the results (\ref{Df fp d = 4}, \ref{Df fp d = 6}) are interesting in their own sense. In $d = 4 - \e$ the two defects that we started with have fused into a different kind of defect. This is not the case in $d = 6 - \e$, and the fused defect is the same as the original defects. In particular this means that if we are given a scalar Wilson defect \eqref{Scalar Wilson loop} in $d = 6 - \e$ (at the conformal f.p.), we cannot determine whether its actually a product of fusion of two such defects or not. This might sound exotic, but it makes sense from an OPE point of view
\begin{equation}
\begin{aligned}
D(-R) D(+R) = D(0) \ .
\end{aligned}
\end{equation}
This serves as a motivation to study fusion of more defects in $d = 6 - \e$. A starting point could be to consider three scalar Wilson defects \eqref{Scalar Wilson loop}, and study whether first fusing two of them and then fuse the resulting fused defect with the last one gives the same result as fusing all three defects at once.

Note that we have not used the conformal symmetry in any way when fusing the defects. Meaning the methods we have used should be applicable to several other kinds of defects and theories as well, assuming we have a Lagrangian description for the defects. E.g. it would be interesting to study fusion of scalar Wilson defects \eqref{Scalar Wilson loop} in theories with other bulk interactions. Or for that matter push our calculations to higher orders in the bulk couplings. Due to how the $Z$-factor for the external field is introduced \cite{Pannell:2023pwz} (see the discussion above \eqref{renorm s}), it could be that the double scaling limit results (upto $\mco(g^4)$, where all bulk-loops are suppressed) of \cite{Bolla:2023zny} are fully valid. 

Another interesting direction to pursue would be to find a more non-trivial fusion (using the methods of this paper), where two defects are fused into a sum of several fused defects
\begin{equation}
\begin{aligned}
D_1D_2 = \sum_{n}C^{12}{}_nD_n \ ,
\end{aligned}
\end{equation}
where the $C^{12}{}_n$ are some kind of OPE coefficients as in a fusion category \cite{etingof2005fusion, bartels2019fusion, douglas2020dualizable}. A possible candidate could be the \textit{monodromy twist defects} (or symmetry defects) \cite{Billo:2013jda, Gaiotto:2013nva, Soderberg:2017oaa, Giombi:2021uae, Gimenez-Grau:2021wiv}. These are of codimension two (exactly), and carry a monodromy constraint for the bulk fields which breaks the global symmetry of the theory. It might be worthwhile to study whether this symmetry breaking can be seen from fusion. 

A generalization to monodromy twist defects are \textit{replica twist defects} (or r\'enyi defects) \cite{SoderbergRousu:2023pbe}. These are used in quantum information to find entanglement entropies \cite{Calabrese:2004eu, Casini:2009sr}. We believe the $c$-function (monotonous under the RG flow) in quantum information can be found from fusion of two replica twist defects \cite{Casini:2015woa}. It would be interesting whether we can apply the techniques from this paper to get more insight into this problem. A drawback with these codimension two defects (both monodromy and replica twist defects) is that we are not aware of any Lagrangian descriptions for them.

\section*{Acknowledgement}

I would like to express my gratitude to Agnese Bissi, Vladimir Bashmakov, Simon Ekhammar, Pietro Longhi and Diego Rodriguez-Gomez for several enriching discussions on (fusion of) defects. I also thank everyone that went to my public defence of my thesis \cite{SoderbergRousu:2023ucv}, where most results in this paper was first presented. 
This project was funded by Knut and Alice Wallenberg Foundation grant KAW 2021.0170, VR grant 2018-04438 and Olle Engkvists Stiftelse grant 2180108.

\bibliographystyle{utphys}
\footnotesize
\bibliography{References}	
	
\end{document}